\documentclass[iop,apj,numberedappendix,twocolappendix,revtex4]{emulateapj}
\usepackage{setspace}
\usepackage{amsfonts,amsmath,amssymb}
\usepackage{txfonts}
\usepackage{graphicx}
\usepackage{bm}
\usepackage{verbatim}
\usepackage{units}
\usepackage[T1]{fontenc}
\usepackage[latin9]{inputenc}
\usepackage{amsbsy}
\usepackage{mathrsfs}
\usepackage{esint}

\begin{document}

\title{Coherent nonhelical shear dynamos driven by magnetic fluctuations at low Reynolds numbers}

\author{J.~Squire\altaffilmark{1} and A.~Bhattacharjee\altaffilmark{1,2}}
\affil{Department of Astrophysical Sciences and Princeton Plasma Physics Laboratory, Princeton University, Princeton, NJ 08543}

\altaffiltext{1}{Max Planck/Princeton Center for Plasma Physics, Department of Astrophysical Sciences and Princeton Plasma Physics Laboratory, Princeton University, Princeton, NJ 08543, USA}
\altaffiltext{2}{Princeton Center for Heliospheric Physics, Department of Astrophysical Sciences and Princeton Plasma Physics Laboratory, Princeton University, Princeton, NJ 08543, USA}

\begin{abstract}
Nonhelical shear dynamos are studied with a particular focus on the possibility of coherent dynamo action. 
The primary results --
serving as a follow up to the results of Squire \& Bhattacharjee [arXiv:1506.04109 (2015)] -- pertain to the 
``magnetic shear-current effect'' as a viable mechanism to drive large-scale magnetic field generation. This effect
raises the interesting possibility that  the saturated state of the small-scale 
dynamo could \emph{drive} large-scale dynamo action, and is likely to be important
in the unstratified regions of accretion disk turbulence. 
In this paper, the effect is studied at low Reynolds numbers, removing the complications of 
small-scale dynamo excitation and aiding analysis by enabling the use of quasi-linear
statistical simulation methods. In addition to the magnetically driven dynamo, new results on 
the kinematic nonhelical shear dynamo are presented. These illustrate
the relationship between coherent and incoherent driving   in such 
dynamos, demonstrating the importance of rotation in determining the relative dominance
of each  mechanism.
\end{abstract}

\section{Introduction}

Understanding the origin and sustenance of astrophysical magnetic
fields remains an outstanding theoretical challenge. Turbulent dynamo,
in which chaotic fluid motions act to amplify or maintain a magnetic
field against dissipation, seems a likely explanation, but many questions
about such processes remain. Interestingly, magnetic fields are generically
observed to be correlated over larger scales than the underlying fluid
motion, and much of dynamo theory has focused on these ``large-scale''
dynamos. The well known $\alpha$-effect \citep{Krause:1980vr} may
explain such behavior, but requires some breakage of symmetry in the
underlying turbulence (e.g., net fluid helicity). In addition, while
the linear (kinematic) regime of such dynamos may be well understood,
there are still significant difficulties regarding dynamo saturation (see \citealt{Brandenburg:2012ct} and
references therein)
including whether it is even possible for large scale fields to grow
to observed amplitudes -- the problem of $\alpha$-quenching \citep{Kulsrud:1992ej,Gruzinov:1994ea,Bhattacharjee:1995ip,Boldyrev:2005ix,Cattaneo:2009cx}. Large scale velocity shear
-- ubiquitous in astrophysical systems due to gravitational forces
-- may have some very important role to play. Most obviously, shear
affects the dynamo through simple stretching of the mean field (the
$\Omega$ effect), but a variety of other more
subtle effects may also enhance dynamo action in various ways (e.g.,
\citealt{Vishniac:2001wo,Blackman:2002cr,Brandenburg:2008kn,Tobias:2014ek}).
In addition, shear seems to allow the growth of large scale dynamos
\emph{without} net helicity or inhomogeneity in the turbulence \citep{Brandenburg:2008bc,Yousef:2008ie}.
Such dynamos may play a fundamental role in a variety of astrophysical
processes where a high degree of symmetry is present, for instance,
the mid-plane of ionized accretion disks.

These non-helical shear dynamos have been an object of fascination
in the dynamo literature for some years. Two fundamentally different
explanations have been proposed for how large-scale fields can be
generated without any symmetry breaking process. The first -- the
so-called ``shear-current effect'' -- is in essence an off-diagonal
turbulent resistivity \citep{Rogachevskii:2003gg,Rogachevskii:2004cx}.
When coupled with the shear, even rather small values of this transport
coefficient can overcome the standard (diagonal) turbulent resistivity
and cause growth of a mean-field dynamo. The second explanation --
the stochastic-$\alpha$ effect -- relies on the idea that even if
the mean $\alpha$-coefficients vanish, sufficiently strong fluctuations
can lead to mean-field growth \citep{Vishniac:1997jo,Silantev:2000tb,Heinemann:2011gt}.
This dynamo is not mean-field in the usual sense since it relies on
the finite size of the system to cause mean-field growth; nonetheless,
given that the universe is sampling a single realization of turbulence,
\emph{not} the ensemble average, such effects could be entirely physical.
At the present time, much of the community appears to have converged
on the idea that non-helical shear dynamos are incoherent in nature;
i.e., the stochastic-$\alpha$ effect is more important than the shear-current
effect. The primary reasoning is that the crucial transport coefficient
required for the shear-current effect appears to have the incorrect
sign, at least at moderate Reynolds numbers \citep{Radler:2006hy,Rudiger:2006gx,Brandenburg:2008bc}.
At the same time, given the variety of different, but related,
incoherent dynamo mechanisms that have been considered \citep{Silantev:2000tb,Heinemann:2011gt,Mitra:2012iv,Richardson:2012gf,Sridhar:2014he},
it seems likely that such effects could be relatively generic. 

Here, following up on a recent paper \citep{HighRm}, we consider the possibility
of large-scale coherent non-helical shear dynamos in the regime of low Reynolds numbers. 
We propose a fundamentally
different mechanism to those discussed above -- that a coherent large-scale
magnetic field can be excited by \emph{small-scale magnetic fluctuations}.
Why should this be important? In any magnetohydrodynamic (MHD) system
above low magnetic Reynolds number, the dynamo at smallest scales
in the turbulence grow the fastest due to the small-scale dynamo
\citep{Schekochihin:2007fy}. Such growth is sufficiently rapid that
it always overwhelms the large-scale field growth and thus a large-scale
field must be able to grow on top of both velocity and magnetic fluctuations
\citep{Kulsrud:1992ej,Boldyrev:2005ix,Cattaneo:2009cx}. This idea
is at the heart of $\alpha$-quenching, where the small-scale magnetic
fluctuations quench the growth of the large-scale field before it
has a chance to reach significant amplitude. Our proposal is that
for non-helical shear-dynamos, the effect of the small-scale
dynamo is \emph{positive}, enhancing the large scale
dynamo growth rate. 

In this paper we focus on understanding such a magnetic
dynamo in the low Reynolds number regime. In this regime the problem becomes substantially
simpler, due to the greater applicability of quasi-linear approximations
and the lack of a small-scale dynamo \citep{Yousef:2008ix}. This
enables the effects of velocity and magnetic fluctuations to be studied
separately (e.g., through driving the induction equation), as well
as allowing simple calculation of transport coefficients and fluctuation
statistics. We see that with sufficiently strong small-scale magnetic
fluctuations, the character of the observed large-scale dynamo changes,
becoming more coherent in time and saturating at higher field strengths. 
That this is a coherent dynamo effect is confirmed through numerical evaluation of the
relevant transport coefficients.
In a recent paper \citep{HighRm}, we have considered the more
relevant case where the magnetic fluctuations are self-consistently
excited by the small-scale dynamo at higher Reynolds numbers, driving a large-scale dynamo once
they reach saturation. 

In addition to studying the magnetic dynamo, we re-examine the kinematic
dynamos presented in \citet{Yousef:2008ix,Yousef:2008ie}, since it
is necessary to understand the intricacies of the kinematic dynamo
before moving on to the magnetically driven case. We find that the
dynamo seen by \citet{Yousef:2008ix} is indeed a stochastic-$\alpha$
effect, of the type suggested by \citet{Heinemann:2011gt}, in their
\emph{non-rotating }examples\emph{. }However, anti-cyclonic (e.g.,
Keplerian) rotation can substantially alter the picture, causing a
coherent dynamo to become possible by changing the sign of the off-diagonal
resistivity. This behavior is well explained by the $\bm{\Omega}\times\bm{J}$,
or Rädler, effect \citep{Krause:1980vr,Moffatt:1982ta}. Although
not commented on by \citet{Yousef:2008ix}, our conclusions are entirely
compatible with their results, nicely explaining the observed trends
in growth rates. 

One of our primary motivations in this work has been improving understanding
of the dynamo observed in zero-net flux magnetorotational (MRI) turbulence
simulations \citep{Brandenburg:1995dc,Hawley:1996gh,Lesur:2008cv}.
Given that such turbulence is simply a shear flow with the addition
of Keplerian rotation, the results presented here
should be applicable to some degree. Of course, self-sustaining MRI
turbulence is highly nonlinear and linear dynamo results will be generally
inapplicable. Instead, one can consider the presence of a large-scale
dynamo instability as an indication that the turbulence will always
be accompanied by large scale structures. Given that MRI turbulence
is both rotating and has strong magnetic fluctuations, it seems reasonable
to surmise from the conclusions reached in this paper that a 
coherent dynamo plays an important role.
Furthermore, our recent statistical calculations (see Sec.~\ref{sec:Equations etc.}) of the nonlinear saturation 
of unstratified MRI turbulence have shown nice agreement with
aspects of self-sustained nonlinear simulations, in particular regarding the dependence
on $\mathrm{Pm}$ \citep{Squire:2015fk}. Since a coherent dynamo is the \emph{only} possible
mechanism in such calculations, this provides strong evidence 
to support  the relevance of the magnetically driven shear dynamo to MRI turbulence.
Our results regarding the MRI dynamo mechanism are
qualitatively consistent with previous computational and analytic studies
 \citep{Lesur:2008fn,Lesur:2008cv}. For the purposes
of understanding MRI turbulence, the nonlinear behavior of the dynamo
will be  important but we leave this complex topic to future work
\citep{Rogachevskii:2004cx,Lesur:2008fn}.

\subsection{Outline}

Since results on both the magnetically driven and kinematic dynamo are presented, 
we feel it helpful to provide a ``roadmap'' for paper's structure.
This is intended to outline how the central results relate to each other,
and convey our motivations for structuring the paper as follows.  

As discussed, the most important results
of this paper are those regarding the ``magnetic shear-current effect,''
which act as a follow up to \citet{HighRm} in the simpler low $\mathrm{Rm}$ regime.
However, to be able to convincingly interpret results -- in particular
observations of magnetic dynamo in nonlinear simulation -- it is necessary 
to first explore the kinematic dynamo, its primary driving mechanisms, and its
dependence on physical parameters. 
Thus, we first present results (Sec.~\ref{Kinematic}) on the dynamo mechanism 
in the simulations of \citet{Yousef:2008ix,Yousef:2008ie}, which 
show that the this kinematic dynamo is primarily incoherent 
(although coherent effects become important with rotation)
and provides a comparison point for later  results on the magnetic dynamo. 
This section also acts to illustrate the effectiveness of the quasi-linear
and statistical simulation methods in disentangling  incoherent
and coherent dynamo mechanisms, and demonstrates 
that the direct measurement of transport coefficients yields results
in agreement with other methods.

The magnetic shear-current effect dynamo is then 
studied in Sec.~\ref{Magnetic}. To argue for its existence, we 
use the same tools as for the kinematic dynamo: qualitative examination 
of the dynamo from direct numerical simulation, statistical simulations at the
same physical parameters as in the kinematic case, and direct measurement
of transport coefficients. We hope that together
these methods provide a strong argument for the existence of
the effect and its potential importance in dynamo theory. 

These sections on the kinematic and magnetic shear dynamos 
are preceded by a theoretical discussion of the different dynamo mechanisms that are
possible in this geometry (Sec.~\ref{sec:Shear-dynamos}), and  an explanation of the numerical methods 
(Sec.~\ref{sec:Equations etc.}), including the quasi-linear approximation 
and statistical simulation methods (CE2). 
The primary purpose of the theoretical discussion 
is to explain the differences between incoherent and coherent 
dynamos, and what properties might be used to distinguish these.
A different stochastic dynamo mechanism \citep{Silantev:2000tb,Sridhar:2014he}, based 
on the work of \citet{Kraichnan:1976gi}, is discussed in App.~\ref{app:Sa dynamos},
where we come to the conclusion that this dynamo is unlikely 
to be causing observed field generation due to the effects of off-diagonal 
$\alpha$ fluctuations.
We finish the paper  in Sec.~\ref{Conclusions}  with conclusions, 
including a detailed comparison with previous works,  as well as suggestions for future studies.

Throughout this paper, our nonlinear simulations will
utilize a similar numerical setup to that of \citet{Yousef:2008ix},
with tall boxes ($L_{z}\gg L_{x}=L_{y}$) to enhance scale separation, and relatively small Reynolds
numbers ($\mbox{Re}=\mbox{Rm}=100$) to avoid the complications of the small-scale dynamo.

\section{Shear dynamos} \label{sec:Shear-dynamos}

In this section we conceptually examine the possibilities of incoherent
(stochastic alpha), and coherent (shear-current) dynamos, arising
from nonhelical turbulence in a Cartesian shearing box. Specifically,
we consider an imposed linear velocity shear, $\bm{U}_{0}=-Sx\hat{\bm{y}}$,
and mean-fields are defined by simple averaging over the horizontal
($x$ and $y$)directions. (Note that $S$ is defined with a negative
sign so as to conform to conventions in the literature on astrophysical
shear flows). A more comprehensive exploration of possible dynamo
mechanisms in this geometry can be found in \citet{Mitra:2012iv}.

In the conventional way, we start by defining mean and fluctuating
fields through the relation $\bm{B}_{T}=\overline{\bm{B}_{T}}+\bm{b}=\bm{B}+\bm{b}$,
where $\bm{B}_{T}$ is the full turbulent magnetic field and $\bar{\cdot}$
is the mean-field average (simply a spatial average over $x$ and
$y$). We shall also make use of the ensemble mean, denoted $\left\langle \cdot\right\rangle $,
which is the average over an ensemble of realizations at the same
physical parameters. Averaging the induction equation [see. Eq.~\eqref{eq:MHD}]
leads to the standard mean field dynamo equations for the mean magnetic
field $\bm{B}$ \citep{Moffatt:1978tc,Krause:1980vr}
\begin{equation}
\partial_{t}\bm{B}=\nabla\times\left(\bm{U}_{0}\times\bm{B}\right)+\nabla\times\mathcal{E}+\frac{1}{\mathrm{Rm}}\triangle\bm{B}.\label{eq:genMF}
\end{equation}
Here $\mathcal{E}=\overline{\bm{u}\times\bm{b}}$ is the electromotive
force, assumed to be of the form $\mathcal{E}_{i}=\alpha_{ij}B_{j}+\eta_{ijk}B_{j,k}+\dots$ due to scale
separation, 
and $\mbox{Rm}$ is the magnetic Reynolds number (inverse normalized
resistivity). Due to the fact that $\bm{B}$ is a
function only of $z$, from $\nabla \cdot \bm{B}=0$ we obtain $B_{z}=0$, and there are only 4 non-zero
components of the $\eta_{ijk}$tensor,\footnote{Specifically $\eta_{i3k}=0$ since $B_{z}=0$, and $\eta_{ijk}=0$
if $k\neq3$ since $\partial_{x}\bm{B}=\partial_{y}\bm{B}=0$.%
} see \citet{Brandenburg:2002cia,Radler:2006hy}. Expanding Eq.~\eqref{eq:genMF}
one obtains
\begin{gather}
\partial_{t}B_{x}=-\alpha_{yx}\partial_{z}B_{x}-\alpha_{yy}\partial_{z}B_{y}-\eta_{yx}\partial_{z}^{2}B_{y}+\eta_{yy}\partial_{z}^{2}B_{x}\nonumber \\
\partial_{t}B_{y}=-SB_{x}+\alpha_{xx}\partial_{z}B_{x}+\alpha_{xy}\partial_{z}B_{y}-\eta_{xy}\partial_{z}^{2}B_{x}+\eta_{xx}\partial_{z}^{2}B_{y},\label{eq:SC sa eqs}
\end{gather}
where the $\eta_{ij}$ are defined as the various non-zero components
of $\eta_{ijk}$. At this stage, the $\alpha$ or $\eta$ are not
assumed constant in time, space, or over realizations (i.e., $\alpha_{ij}\neq\left\langle \alpha_{ij}\right\rangle $)
-- indeed with the mean-field average taken over a finite sized domain
they can fluctuate strongly. General symmetry arguments \citep{Radler:2006hy,Brandenburg:2008bc}
show that $\left\langle \alpha_{ij}\right\rangle =0$, while there
are no such constraints on the form of $\eta_{ij}$ when effects that
break the isotropy of the turbulence are present (e.g., shear, rotation).
We shall assume that the diagonal components of the resistivity, $\eta_{yy}$
and $\eta_{xx}$, are positive, since the scale separation assumptions
of mean-field theory will presumably become invalid if this is not
the case. 

The two fundamental dynamo mechanisms we will examine in this work
are:
\begin{description}
\item [{Coherent~shear~dynamo}] This dynamo arises primarily from the
coupling between the off-diagonal resistivity $\eta_{yx}$ and the
shear term $-SB_{x}$. Specifically, for Eq.~\eqref{eq:SC sa eqs}
 with $\alpha_{ij}=0$ and $\eta_{yy}=\eta_{xx}=\eta_{t}$ (equality of the diagonal resistivities is just for simplicity, 
 the dynamo is not changed qualitatively by relaxing this),
it is straightforward to show that an eigenmode with the spatial structure
$B_{i}=B_{i0}e^{ikz}$ has the growth rate 
\begin{equation}
\gamma_{\eta}=k\sqrt{\eta_{yx}\left(-S+k^{2}\eta_{xy}\right)}-k^{2}\eta_{t}.\label{eq:gamma SC}
\end{equation}
Neglecting $\eta_{xy}$ by assuming $|k^{2}\eta_{xy}|\ll S$
for all $k$ for which scale separation  holds, positive
dynamo growth is possible if $-S\eta_{yx}>0$, $k\sqrt{-\eta_{yx}S}>k^{2}\eta_{t}$.
The maximum growth rate is $\gamma_{\eta}=|S\eta_{yx}|/4\eta_{t}$,
obtained at $k=\sqrt{|S\eta_{yx}|}/2\eta_{t}$ (if this
wavenumber fits in the box). For a single mode of this dynamo, $B_{x}$
and $B_{y}$ are $\pi$ out of phase, $B_{x}=-k\sqrt{|\eta_{yx}/S|}B_{y},$
and their phases are constant in time, meaning $Re\langle B_{x}B_{y}^{*}\rangle =\sqrt{\langle B_{x}B_{x}^{*}\rangle \langle B_{y}B_{y}^{*}\rangle }$.
A nonzero $\eta_{yx}$ can arise from the effect of shear, the \emph{shear-current
effect} \citep{Rogachevskii:2003gg}, from rotation \citep{Krause:1980vr,Radler:2003gg},
the \emph{$\bm{\Omega}\times \bm{J}$ (or Rädler) effect, }or from a combination
of both. Since with the shear-current effect, $\eta_{yx}\propto S$,
the maximum growth rate of the coherent dynamo should scale as $\gamma\sim S^{2}$ (this also holds with rotation if
$\Omega/S$ is fixed, e.g., Keplerian rotation).
\item [{Stochastic~alpha~effect}] This dynamo arises from the combination
of zero-mean $\alpha_{yy}$ fluctuations and the mean shear $S$.
Consider Eq.~\eqref{eq:SC sa eqs} with $\eta_{xy}=\eta_{yx}=0$, $\langle \alpha_{ij}\rangle =0$,
and again take $\eta_{yy}=\eta_{xx}=\eta_{t}$. For simplicity,\footnote{When the shear is larger than fluctuations in $\alpha$, $\left\{ \alpha_{xy},\alpha_{yx},\alpha_{xx}\right\} $
are each subdominant to $\alpha_{yy}$ in their effect on the growth
rate, see \citet{Mitra:2012iv}.%
} we set $\alpha_{xy}(t)=\alpha_{yx}(t)=\alpha_{xx}(t)=0$,
assume white noise fluctuations in $\alpha_{yy}$, $\langle \alpha_{yy}\left(t\right)\alpha_{yy}\left(t'\right)\rangle =\langle \alpha_{yy}^{2}\rangle \delta\left(t-t'\right)$,
and again take $B_{i}=B_{i0}e^{ikz}$. One can show using standard
techniques \citep{Vishniac:1997jo} that while $\langle B_{i}\left(t\right)\rangle$ decays due to turbulent resistivity,
it is possible for $\langle B_{i}B_{j}^{*}\rangle $ to
grow at the rate 
\begin{equation}
\gamma_{\alpha}=\left(\frac{k^{2}S^{2}\left\langle \alpha_{yy}^{2}\right\rangle }{2}\right)^{1/3}-k^{2}\eta_{t}.\label{eq:gamma SA}
\end{equation}
Thus, positive dynamo growth is possible if fluctuations in $\alpha$
are sufficiently large. The maximum growth rate of this dynamo is
$\gamma=0.074\, S\sqrt{D_{yy}/\eta_{t}}$ , obtained at $k=\sqrt{S}(D_{yy}/54\eta_{t}^{3})^{1/4}$.
Note that in any single realization of this dynamo, as observed in
simulation, $B_{x}$ and $B_{y}$ will grow approximately exponentially;
the fact that $\left\langle B_{i}\right\rangle =0$ would only become
apparent if a large ensemble of simulations were carried out at the
same physical parameters (with the same initial conditions for the
mean field). Importantly, initial conditions must 
be forgotten over the timescale associated with the turbulent resistivity, $t\sim (k^{2}\eta_{t})^{-1}$
(since $\left\langle B_{i}(t)\right\rangle$ simply decays exponentially), which 
implies that the dynamo cannot have a constant phase as it grows in time. 
For a single mode of the dynamo, $B_{x}$ and $B_{y}$
are on average $\pi/4$ out of phase (as for the coherent $\alpha$
shear dynamo), $Re\langle B_{x}B_{y}^{*}\rangle =2^{-1/2}\sqrt{\langle B_{x}B_{x}^{*}\rangle \langle B_{y}B_{y}^{*}\rangle }$. The stochastic-$\alpha$
dynamo will also have a dependence on the horizontal domain size,
since averaging over a larger domain will decrease the size of the
fluctuations in $\alpha$, thus decreasing the magnitude of the growth
rate. More information, including the effects of other nonzero $\alpha$
coefficients and correlations between different $\alpha_{ij}$, can
be found in \citet{Mitra:2012iv}. The stochastic alpha dynamo has
also been derived from the MHD equations directly by quasi-linearly considering
a collection of forced shearing waves \citep{Heinemann:2011gt,McWilliams:2012du}. A fundamentally different type of stochastic-$\alpha$ shear dynamo 
has also been proposed and studied in \citet{Silantev:2000tb,Sridhar:2014he}. 
We explore this further in App.~\ref{app:Sa dynamos}, where we arrive at the
conclusion that the effect is unlikely to be driving the large scale dynamos studied in this manuscript
due to the adverse effect of off-diagonal $\alpha$ fluctuations.
\end{description}
Of course, in a real turbulent situation, these two dynamos can be
mixed, and distinguishing the two may be rather difficult. In particular,
the $\eta_{ij}$ coefficients discussed for the coherent shear dynamo
will also fluctuate in time and the mean fields will generally be
noisy, even if the stochastic alpha effect is not the dominant dynamo
driver. In this work we will use a variety of methods to compare the
two in different physical situations, from directly calculating transport
coefficients, to simply observing mean-field temporal evolution. 

It is interesting to note that the growth rate of a stochastic-$\alpha$ dynamo
can be arbitrarily increased or decreased by changing the volume of the mean-field average. 
In particular,  an increase in the
volume of the average by a factor $a$ must lead to a reduction in the magnitude 
of $\langle \alpha^{2} \rangle$ by $a$ also, assuming the turbulence in  each sub-volume is statistically independent. 
With smaller $\langle \alpha^{2} \rangle$, a reduction in the dynamo growth rate would result. In fact, we see this effect explicitly
in the simulations presented in Sec.~\ref{Kinematic} by simply doubling the horizontal dimensions
of our domain, keeping all other parameters fixed. 

\section{Equations and numerical method}\label{sec:Equations etc.}

In this section we outline the equations solved, as well as outlining
our quasi-linear and statistical methods. The fundamental equations
are the nonlinear magnetohydrodynamic equations with a background
shear flow $\bm{U}_{0}=-Sx\hat{\bm{y}}$ and possible rotation
\begin{align}
\frac{\partial\bm{U}_{T}}{\partial t} & -Sx\frac{\partial\bm{U}_{T}}{\partial y}+\left(\bm{U}_{T}\cdot\nabla\right)\bm{U}_{T}+2\Omega\bm{\hat{z}}\times\bm{U}_{T}+\nabla p=\nonumber \\
 & \qquad SU_{Tx}\bm{\hat{y}}+\bm{B}_{T}\cdot\nabla\bm{B}_{T}+\bar{\nu}\nabla^{2}\bm{U}_{T}+\bm{\sigma}_{\bm{u}},\nonumber \\
\frac{\partial\bm{B}_{T}}{\partial t} & -Sx\frac{\partial\bm{B}_{T}}{\partial y}=-SB_{Tx}\bm{\hat{y}}+\nabla\times\left(\bm{U}_{T}\times\bm{B}_{t}\right)+\bar{\eta}\nabla^{2}\bm{B}_{t}+\bm{\sigma}_{\bm{b}},\nonumber \\
 & \nabla\cdot\bm{U}_{T}=0,\;\;\;\nabla\cdot\bm{B}_{t}=0.\label{eq:MHD}
\end{align}
Here $\Omega$ is a mean rotation of the frame, and $\bar{\nu}$ and
$\bar{\eta}$ are the normalized viscosity and resistivity respectively.
Since all quantities are normalized to one it is convenient to define
$\mbox{Re}=1/\bar{\nu}$ and $\mbox{Rm}=1/\bar{\eta}$ for the Reynolds
and magnetic Reynolds number. The driving noise ($\bm{\sigma}_{\bm{u}}$ and $\bm{\sigma}_{\bm{b}}$)
is non-helical and white in time, localized in wavenumber
around $k=6\pi$ with width $6\pi/5$,
and is used to generate an homogenous bath of small scale velocity
and/or magnetic fluctuations,\footnote{This forcing is of the same form as \citet{Yousef:2008ix}.}
$\bm{U}_{T}$ and $\bm{B}_{T}$ in Eq.~\eqref{eq:MHD} denote the
full turbulent fields ($\bm{U}_{T}$
is the velocity not including the background shear) \textendash{}
while this notation may seem cumbersome, for the remainder of the
article we will split $\overline{\bm{B}_{T}}$ and $\overline{\bm{U}_{T}}$
into their mean ($\bm{B}=\overline{\bm{B}_{T}}$, $\bm{U}=\overline{\bm{U}_{T}}$)
and fluctuations ($\bm{u}$, $\bm{b}$). We have deliberately not
normalized Eq.~\eqref{eq:SC sa eqs} with respect to the rotation
$\Omega$ as is standard in MRI studies \citep{Balbus:1998tw}, so
as to allow study of shear without rotation. Throughout this work
we consider initially homogenous turbulence with zero average helicity.
We use a Cartesian box of dimensions $\left(L_{x},L_{y},L_{z}\right)$ with
periodic boundary conditions in $z$ and $y$, and shearing periodic
boundary conditions in $x$. 

Our primary tool for solving Eq.~\eqref{eq:MHD} is the SNOOPY code
\citep{Lesur:2007bh}. This solves Eq.~\eqref{eq:MHD} with a Fourier
pseudo-spectral method in the shearing frame, using standard methods
for dealiasing and remapping. Our standard simulation setup is to
seed from random Gaussian initial conditions in $\bm{u}$ and $\bm{B}$
at a very small amplitude and reasonably large scales (wavelengths
greater than $\sim0.2$). The forcing, $\bm{\sigma}_{u}$ (and sometimes
$\bm{\sigma}_{\bm{b}}$), causes a small scale turbulent bath of fluctuations,
and we study growth of the dynamo on larger scales than the forcing
(i.e., $k<15$). As in \citet{Yousef:2008ix}, the separation of scales
between mean-fields and fluctuations is aided by choosing a box that
is very elongated in the $z$ direction, $L_{z}>L_{x},L_{y}$, We
study the development of the dynamo by numerically averaging $\bm{B}$
over $x$ and $y$ to obtain the mean magnetic fields, $\overline{\bm{B}}$,
see Sec.~\ref{sec:Shear-dynamos}. Overall, the numerical setup of
our nonlinear runs is nearly identical to that of \citet{Yousef:2008ix},
aside from the addition of forcing in the induction equations in some
simulations. The Reynolds numbers as defined are with respect to the
large scale shear. It is also useful to keep in mind more standard
definitions of these using the small-scale velocity, denoted $\mathrm{Rm}_{f}$
and $\mathrm{Re}_{f}$. Since we use the same forcing spectrum throughout
this work, these are related to Rm and Re through
\[
\mathrm{Rm}_{f}=\frac{u_{rms}}{k_{f}}\mathrm{Rm}=0.053u_{rms}\mathrm{Rm}
\]
with the similar definition for $\mathrm{Re}_{f}$. Most of the calculations
presented in this work have $\mathrm{Re}_{f}=\mathrm{Rm}_{f}\approx5$.

\subsection{Quasi-linear method and statistical simulation}

For certain aspects of this study, we have found it to be very useful
to study the dynamo using a quasi-linear model and statistical simulation
in addition to the nonlinear MHD equations. Here we outline these
methods and the motivation behind them. More details can be found
in \citet{Farrell:2014we,Squire:2015fk}.

The basic idea of the quasi-linear model is to split the mean field
and fluctuations \emph{before} solving the equations, neglecting nonlinearities
in the fluctuation equations. The equations are thus easily derived
by substitution of $\bm{U}_{T}=\bm{U}+\bm{u}$, $\bm{B}_{T}=\bm{B}+\bm{b}$
into Eq.~\eqref{eq:MHD}, followed by a split of each equation into
a mean and fluctuating part. This leads to
\begin{align}
\left(\partial_{t}-Sx\partial_{y}\right)\bm{U} & =-2\Omega\hat{\bm{z}}\times\bm{U}+SU_{x}\hat{\bm{y}}+\nonumber \\
\bar{\nu}\partial_{z}^{2}\bm{U} & +\left(-\overline{\bm{u}\cdot\nabla\bm{u}}+\overline{\bm{b}\cdot\nabla\bm{b}}\right),\nonumber \\
\left(\partial_{t}-Sx\partial_{y}\right)\bm{B} & -SB_{x}\hat{\bm{y}}+\bar{\eta}\partial_{z}^{2}\bm{B}+\nabla\times\overline{\left(\bm{u}\times\bm{b}\right)},\nonumber \\
\partial_{z}B_{z} & =\partial_{z}U_{z}=0\label{eq:QL MF eqns}
\end{align}
for the mean-fields, and\begin{subequations} \label{eq:QL fluct eqs}
\begin{align}
\left(\partial_{t}-Sx\partial_{y}\right)\bm{u} & =-2\Omega\bm{\hat{z}}\times\bm{u}+Su_{x}\hat{\bm{y}}+\bar{\nu}\nabla^{2}\bm{u}-\nabla p-\nonumber \\
\left(\bm{u}\cdot\nabla\bm{U}\right. & \left.+\bm{U}\cdot\nabla\bm{u}\right)+\left(\bm{B}\cdot\nabla\bm{b}+\bm{b}\cdot\nabla\bm{B}\right)+\bm{\sigma}_{\bm{u}},\label{eq:QL fluct eqs u} \\
\left(\partial_{t}-Sx\partial_{y}\right)\bm{b} & =Sb_{x}\hat{\bm{y}}+\bar{\eta}\nabla^{2}\bm{B}_{t}+\nabla\times\left(\bm{u}\times\bm{B}\right)\nonumber \\
 & \nabla\times\left(\bm{U}\times\bm{b}\right)+\bm{\sigma}_{\bm{b}},\label{eq:QL fluct eqs b}\\
\nabla\cdot\bm{u}=0, & \;\;\;\nabla\cdot\bm{b}=0.\nonumber
\end{align}
\end{subequations}for the fluctuating fields. Note that only the
fluctuations are driven by noise. In some cases, it is also convenient
to not include the nonlinear stress feedback on the mean-fields, keeping
these static in time by simply setting $\partial_{t}\bm{U}=\partial_{t}\bm{B}=0$.
This allows the calculation of the transport coefficients directly [e.g., $\eta_{yx}$,
or $\alpha_{yy}\left(t\right)$ in Eq.~\eqref{eq:genMF}] 
for the chosen form of $B_{x}$, $B_{y}$, or even $\bm{U}$, and
is essentially a quasi-linear test field method.

\paragraph{Statistical simulation}

Noticing that Eqs.~\eqref{eq:QL fluct eqs} are linear and driven by
white noise, we can solve for its statistics directly. This method
has been termed Stochastic Structural Stability Theory (S3T) \citep{Farrell:2003ud}
or the Second Order Cumulant Expansion (CE2) \citep{Marston:2008gx}
\textendash{} we will use the term CE2 in this work. In the context
of Eqs.~\eqref{eq:QL MF eqns} and \eqref{eq:QL fluct eqs}, we form
the equation for the second order statistics of $\bm{u}$ and $\bm{b}$,
\[
\mathcal{C}=\left(\begin{array}{cc}
\langle \bm{u}\bm{u}^{\dagger}\rangle  & \langle \bm{u}\bm{b}^{\dagger}\rangle \\
\langle \bm{b}\bm{u}^{\dagger}\rangle  & \langle \bm{b}\bm{b}^{\dagger}\rangle 
\end{array}\right).
\]
This is 
\begin{equation}
\partial_{t}\mathcal{C}=\mathcal{A}\mathcal{C}+\mathcal{C}\mathcal{A}^{\dagger}+\mathcal{Q},\label{eq:C eq}
\end{equation}
where $\left\langle \bm{\sigma}\left(t\right)\bm{\sigma}\left(t'\right)\right\rangle =\delta\left(t-t'\right)\mathcal{Q}$ 
(with $\bm{\sigma}$ representing both the $\bm{u}$ and $\bm{b}$
noise) and $\mathcal{A}\left(\bm{U},\bm{B}\right)$ is the linear
operator in Eq.~\eqref{eq:QL fluct eqs}; i.e., Eq.~\eqref{eq:QL fluct eqs} is equivalent to
\[
\partial_{t}\left(\begin{array}{c}
\bm{u}\\
\bm{b}
\end{array}\right)=\mathcal{A}\left(\begin{array}{c}
\bm{u}\\
\bm{b}
\end{array}\right)+\left(\begin{array}{c}
\bm{\sigma}_{\bm{u}}\\
\bm{\sigma}_{\bm{b}}
\end{array}\right).
\]
If we then set $\overline{f\left(\bm{u},\bm{b}\right)}=\left\langle \overline{f\left(\bm{u},\bm{b}\right)}\right\rangle $
in Eq.~\eqref{eq:QL MF eqns}, we can drive the mean-fields with
the \emph{deterministic }nonlinear stresses obtained through simultaneous
solution of Eq.~\eqref{eq:C eq}. Note that equating $\overline{f\left(\bm{u},\bm{b}\right)}$
with $\left\langle \overline{f\left(\bm{u},\bm{b}\right)}\right\rangle $
is \emph{not }valid due to the finite size of our system and the fluctuating
nature of horizontal averaged quantities (see Sec.~\ref{sec:Shear-dynamos}),
we discuss this in more detail below.

We use a Fourier pseudo-spectral numerical method for both direct
quasi-linear simulation (DQLS) {[}Eqs.~\eqref{eq:QL fluct eqs} and
\eqref{eq:QL MF eqns}{]} and solving the the CE2 equations {[}Eqs.~\eqref{eq:C eq}
and \eqref{eq:QL MF eqns}{]}. The codes are written in \texttt{c++} with MPI
parallelization, and use $3/2$ dealiasing with the remapping method
of \citet{Lithwick:2007ge}. Since one solves for the inhomogenous
fluctuation statistics in $z$, the CE2 code requires a grid of size
$N_{x}\times N_{y}\times\left(mN_{z}\right)^{2}$ (where $m$ is the
number of variables), and CE2 calculations can be relatively expensive.
The codes have been verified and tested in a variety of ways; see
\citet{Squire:2015fk} for more information.

\paragraph{Discussion of the quasi-linear method}

The quasi-linear model involves a rather drastic approximation to
the full nonlinear equations. What do we gain by studying such a system?
Generally, such models have allowed a simpler interpretation and study of
large scale structure growth in turbulence, and been rather useful
in a variety of geophysical, plasma, and fluid dynamics problems (see, for example,
\citealt{Farrell:2009ke,Farrell:2012jm,Tobias:2013hk,Parker:2013hy}). In previous
work \citep{Squire:2015fk} we have found surprisingly good agreement
between saturated states of CE2 for the MRI system and nonlinear simulation,
in particularly a strong scaling with magnetic Prandtl number. 

In the context of the work presented here, the methods provide a simple
way to calculate transport coefficients by fixing the fields in both the magnetically 
driven and kinematic cases, followed by an unambiguous check
that a mean-field dynamo can be observed at the same physical parameters.
While versions of the (nonlinear) test-field method exist that explicitly take into account magnetic
fluctuations \citep{Rheinhardt:2010do}, these are relatively complicated and in the 
early stages of development. 
However, the most important benefit of the quasi-linear methods is afforded by the comparison between CE2 and DQLS. 
This provides an
unambiguous test of whether the dynamo is coherent or incoherent, since statistical averages
are inserted directly into the CE2 mean-field equations and an incoherent dynamo is not possible. 
Thus, if similar results are observed between CE2 and DQLS, we can be sure 
that the dynamo arises through  $\eta_{ij}$ transport coefficients. Another 
interesting aspect of CE2 is that long periods of exponential growth in the mean-field can 
be observed, even when strong small-scale magnetic fluctuations are present (e.g., due to magnetic
driving). This is in contrast to DNS or DQLS, where it is generically difficult to observe
exponential dynamo growth in the presence of strong magnetic fluctuations, since the finite 
size of the domain causes the mean-field to  come into near equipartition with the fluctuations
almost instantaneously.

Finally, we note that
CE2 calculations in fixed mean fields are in essence the same calculation
as the semi-analytic results presented in \citet{Sridhar:2010it,SINGH:2011ho}.
Our results agree with their findings in the rotation-less case with
only momentum equation forcing.

\section{Kinematic dynamo}\label{Kinematic}

Before exploring the dynamo with magnetic fluctuations it is important
to fully understand the kinematic dynamos presented in \citet{Yousef:2008ix}.
With this aim, we have reproduced many of their simulations 
across a variety of $S$, $\Omega$ and $L_{z}$, to better understand the 
fundamental dynamo mechanisms. We present the
most relevant of these results here. 

For the kinematic dynamo we drive
only the momentum equation in Eq.~\eqref{eq:MHD} (i.e., $\bm{\sigma}_{\bm{b}}=0$),
and at these  Reynolds numbers the  small scale field arises purely from tangling
of the mean-field by velocity, $\nabla\times\left(\bm{u}\times\bm{B}\right)$,
an effect which is quite distinct from the small-scale dynamo \citep{Schekochihin:2007fy}.
In both the rotating and non-rotating cases, we see a mean field dynamo
above some threshold in $L_{z}$ and $\langle \bm{\sigma}_{\bm{u}}^{2}\rangle $
i.e., the dynamo is only excited in a sufficiently tall box if driven hard enough. 
Given the scaling of the growth rates
in Eqs.~\ref{eq:gamma SC} and \ref{eq:gamma SA} and the fact that
$\eta_{t}$ is the sum of a turbulent and physical resistivity, this
behavior is expected for both incoherent and coherent dynamos.\footnote{One 
might expect the dynamo to disappear again if $\langle \bm{\sigma}_{\bm{u}}^{2}\rangle $ is
increased further, due to the increase in $\eta_t$ causing the dynamo to become stable. 
This behavior is seen in the quasi-linear case, but it seems 
that at these parameters in the nonlinear runs, a small-scale dynamo is excited 
before this occurs.}  Our
main finding is that the non-rotating case is a stochastic alpha dynamo
(essentially that explored analytically in \citealt{Heinemann:2011gt}, but including non-zero $k_{z}$)
but that rotation qualitatively changes the mechanism, decreasing $\eta_{yx}$ to negative values
and causing the dynamo to become more coherent. 

In both the non-rotating and rotating cases we run DNS, DQLS, and CE2
calculations at identical parameters. The purpose of this comparison
is primarily to illustrate the difference between CE2 and DQLS (due to the 
incoherent mean-field dynamo), while showing that the nonlinear 
case exhibits a qualitatively similar dynamo to DQLS. Although the
spatiotemporal evolution of the mean-field is similar in each case, we shall see that the DNS and DQLS 
runs exhibit slightly different growth rates. This can be attributed to inaccuracies in the
quasi-linear approximation at these Reynolds numbers.

\paragraph*{Non-rotating dynamo}

\begin{figure}
\begin{centering}
\includegraphics[width=1.0\columnwidth]{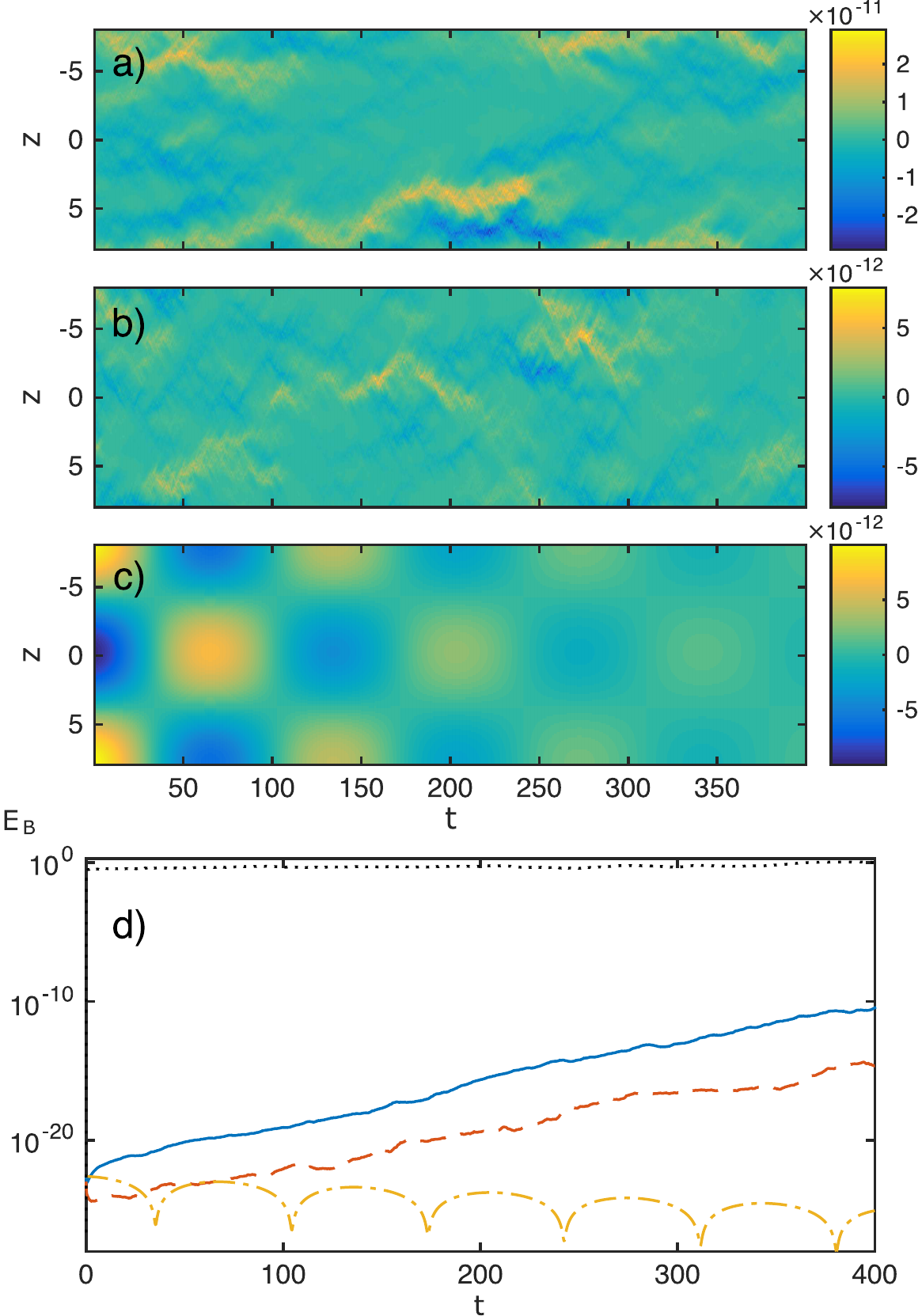}
\caption{(a-c) Illustration of $B_{y}\left(z,t\right)$ from non-rotating turbulence
with $S=2,$ $L_{z}=16$, $u_{rms}=0.8$, for DNS, DQLS, and CE2 from (a)-(c). 
In the two direct runs (a) and (b), we remove the exponential growth (i.e., plot $e^{-\gamma t}B_{y}\left(z,t\right)$ where 
$\gamma$ is the measured growth rate) so that the full time evolution can be observed.
(d) Growth in time of
the mean field for the nonlinear equations (solid,blue), quasi-linear
DNS (dashed,red) and CE2 (dash-dot, yellow), each at the same physical parameters as in
(a). While both nonlinear and quasi-linear DNS exhibit a positive
mean-field dynamo, the CE2 calculation does not, illustrating the
dynamo must be incoherent. The dotted black line shows the energy of $u$ fluctuations. 
\label{fig:Non rotating low Rm}}
\end{centering}
\end{figure}
Fig.~\ref{fig:Non rotating low Rm} illustrates the growth of
the nonrotating dynamo using DNS, DQLS and CE2, at $S=2$  and  $L_{z}=16$, $L_{x}=L_{y}=1$. 
As in \citet{Yousef:2008ix}, we use a resolution $\left(32, 32, 512\right)$ for DNS and DQLS, but use 
$\left(32, 32, 256\right)$ for the CE2 run since the algorithm scales with $N_{z}^{2}$ so is quite computationally  expensive.\footnote{We have verified that 
identical results are obtained at half this resolution and are confident that $N_{z}=256$ is sufficient to resolve all important scales.}

\begin{figure}
\begin{centering}
\includegraphics[width=1.0\columnwidth]{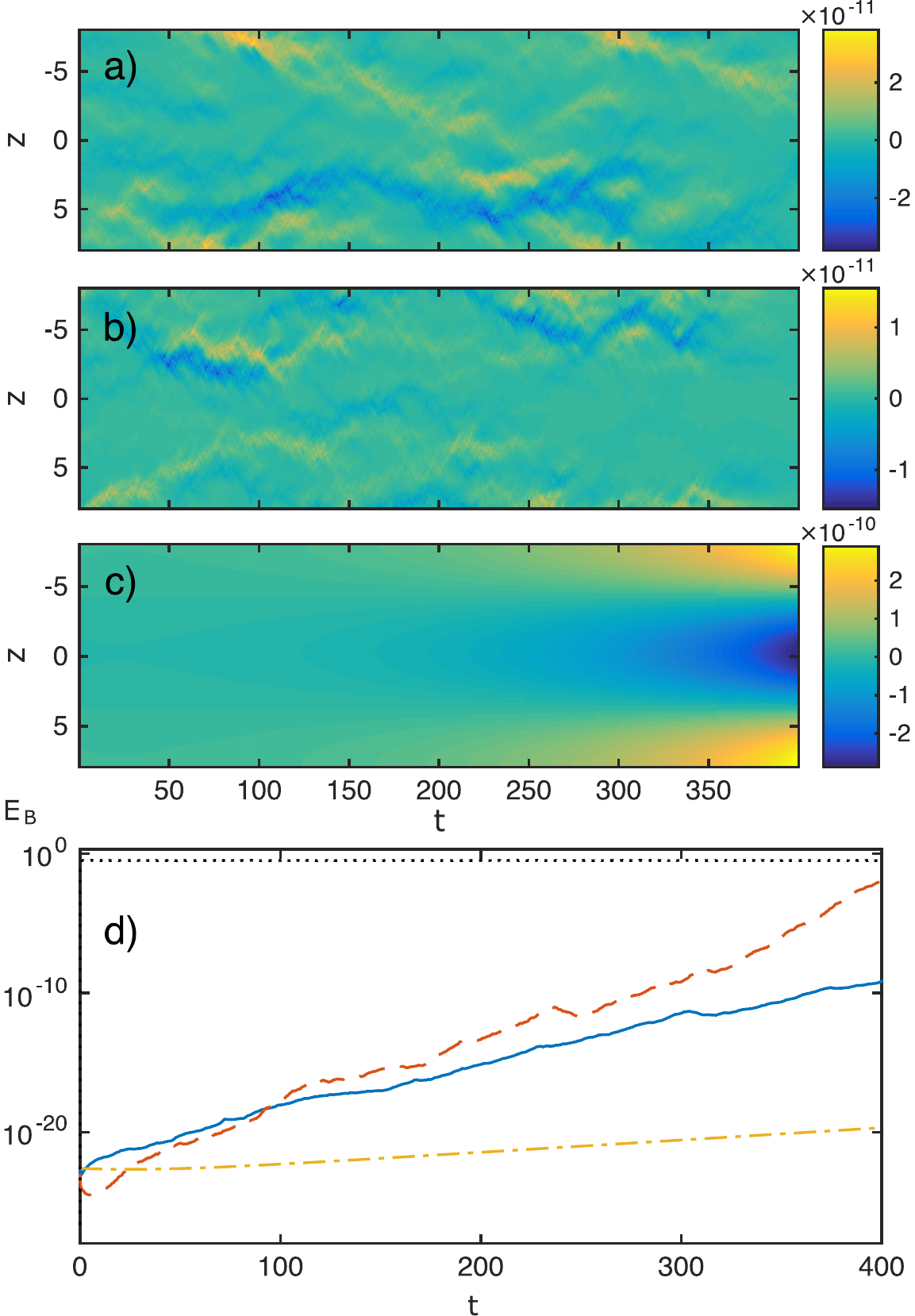}
\caption{Same as Fig.~\ref{fig:Non rotating low Rm} but for 
Keplerian rotating turbulence,  $u_{rms}=0.75$ (velocity flucutations 
are suppressed slightly by the rotation).
In contrast to Fig.~\ref{fig:Non rotating low Rm}, the CE2 calculation
also shows a growing dynamo, albeit at a much smaller growth rate, 
illustrating that the dynamo is partially
coherent. \label{fig:Rotating low Rm}}
\end{centering}
\end{figure}

Firstly, it is worth noting that the mean-field, as plotted in Fig.~\ref{fig:Non rotating low Rm}(a-b), is 
truly a ``large-scale'' dynamo. We can estimate the wavenumber of $B_{y}$ as approximately $3\times2\pi/L_{z}\approx 1.2$, 
far smaller than the forcing scale, $k_{f} = 6\pi$.
Next, let us compare the CE2 with the the nonlinear and quasi-linear
DNS. It is evident that the dynamo in this case is purely incoherent -- while 
slow mean-field growth is observed in DNS and DQLS, the magnetic field simply decays in the CE2 simulation in exactly
the way that would be expected due to a positive $\eta_{yx}$ coefficient. It
is also worth noting the qualitative appearance of the mean-fields, which appear to
wander randomly, as expected due to a stochastic-$\alpha$  effect.
A final piece
of evidence for the incoherency of this non-rotating dynamo comes
from doubling the box size in the $x$ and $y$ dimensions, keeping
all other parameters fixed%
\footnote{We would like to thank A. Schekochichin for suggesting this numerical
experiment.%
} (not shown). This causes the growth rate of the mean-field dynamo
to change from $\gamma=0.062$ [for the dynamo in ~\ref{fig:Non rotating low Rm}(a)]
to being almost stable, $\gamma=0.0096$, and since a coherent dynamo
should be mostly unaffected by such a change (unless the added 
wavenumbers significantly affect the transport coefficients), this constitutes a simple check
of the dynamo's incoherency without using of the quasi-linear
approximation.

\paragraph*{Rotating dynamo}

In Fig.~\ref{fig:Rotating low Rm}, we illustrate the same calculations
as Fig.~\ref{fig:Non rotating low Rm}, but with a Keplerian Coriolis
force {[}$\Omega=2/3S$ in Eq.~\eqref{eq:MHD}{]} added.
While the dynamo in the quasi-linear and nonlinear direct
simulations are similar (with a slightly higher growth rate) to the non-rotating case, the CE2 dynamo
is markedly different, exhibiting mean-field growth. This illustrates
that adding net rotation to the system enabled a coherent
dynamo, which can be understood as arising from a change in sign of
$\eta_{yx}$ (see also Figs.~\ref{fig:kinematic transport Nrot}-\ref{fig:kinematic transport Rot} below). This effect is simply
the well-known Rädler, or $\bm{\Omega}\times\bm{J}$, effect 
\citep{Krause:1980vr,Moffatt:1982ta}. This idea seems to
have been missed in \citet{Yousef:2008ix}, who state ``There does
not appear to be much difference, qualitative or quantitative, between
the rotating and nonrotating cases.'' The finding agrees with analytic results \citep{Analytic},
which show that the contributions to $\eta_{yx}$ from rotation
and the shear have identical forms, and together give $\eta_{yx}\propto S-2\Omega$.
For Keplerian rotation this is slightly negative, leading to the possibility
of coherent dynamo growth. 
Finally, we have again doubled the horizontal dimensions of the box for this rotating case (not shown), which
causes the dynamo growth rate to drop from $\gamma = 0.067$ [in Fig.~\ref{fig:Rotating low Rm}(a)]
to $\gamma = 0.041$. A comparison with the results in the previous paragraph 
($\gamma=0.062$ and $\gamma=0.0096$ in the narrow and wide boxes respectively) shows that although
in the narrow box ($L_{x}=L_{y}=1$) rotation causes only a minor difference to the growth rate 
(because the stochastic-$\alpha$ effect significantly overwhelms the coherent dynamo), in the wider box (where 
fluctuations in $\alpha$ have been significantly reduced),
the difference in growth rates is much more substantial. 
This behavior is consistent with the rotating dynamo being driven through a
combination of stochastic-$\alpha$ and coherent effects [see Eq.~\ref{eq:gamma sa and sc} below],
with the coherent effect being mostly unmodified by the change in box dimensions.

\begin{figure}
\begin{centering}
\includegraphics[width=1.0\columnwidth]{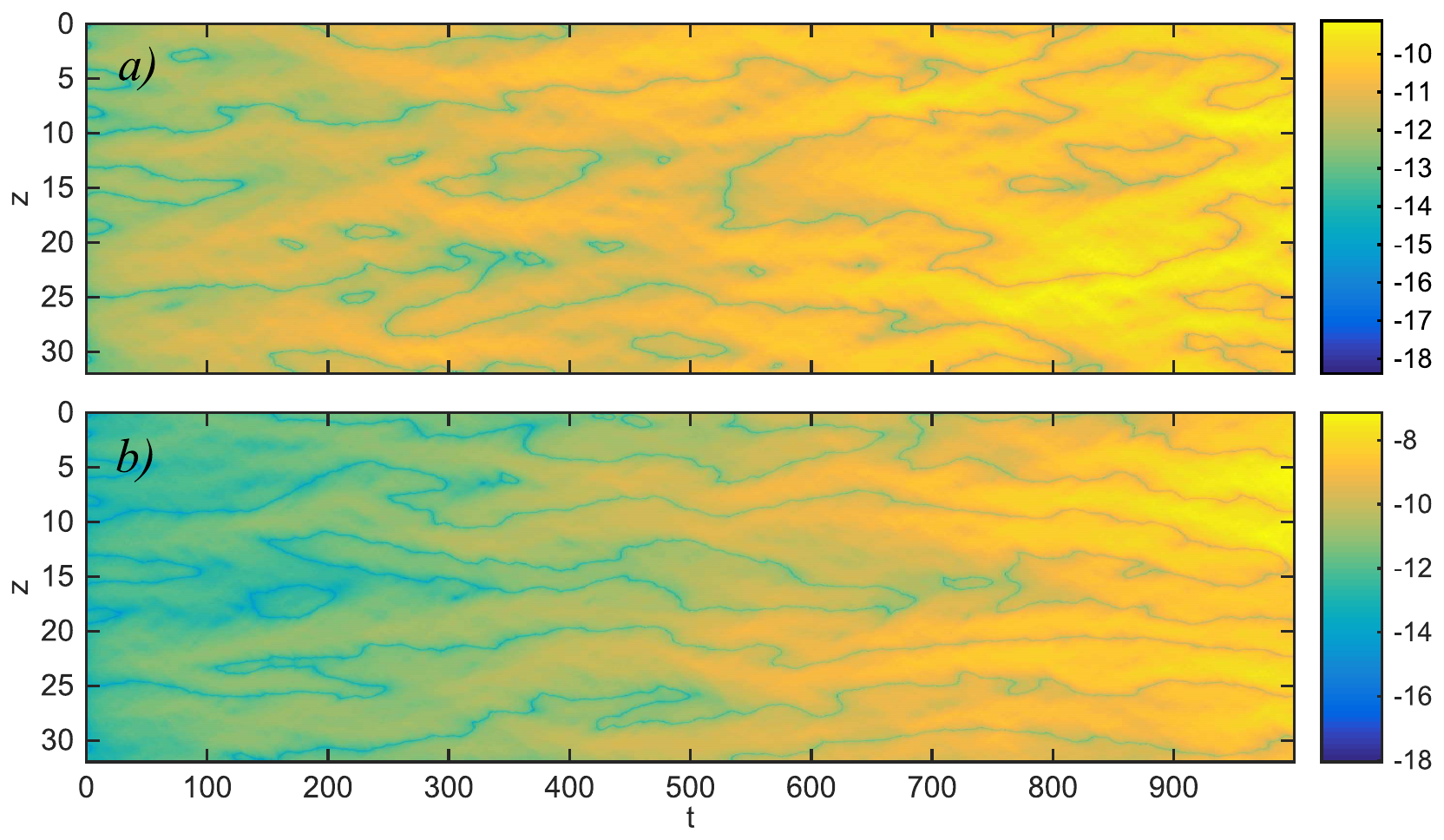}

\caption{Spatiotemporal evolution of $\log_{10}\left(\left|B_{y}\left(z,t\right)\right|\right)$
at $S=1$, $L=32,$ for (a) nonrotating case ($u_{rms} \approx 0.47$), 
(b) Keplerian rotation  ($u_{rms} \approx 0.43$, again velocity fluctuations are slightly suppressed by rotation). 
The $\log$ color scale is chosen so as to easily see the mean-field phase evolution. 
The difference in the two dynamos is evident from the
evolution of the phase of $B_{y}$ as the dynamo grows. While in the
non-rotating case the phase wanders somewhat randomly, as is characteristic
of an incoherent dynamo mechanism (see Sec.~\ref{sec:Shear-dynamos}),
we see a relatively constant phase of $B_{y}$ in the case with rotation.
Note also the faster growth rate of the rotating dynamo. 
\label{fig:logBy}}
\end{centering}
\end{figure}
While the $B_{y}(z,t)$ evolution pictured in Figs.~\ref{fig:Non rotating low Rm}(a) and \ref{fig:Rotating low Rm}(a)
looks qualitatively rather similar between the rotating and non-rotating runs, this is 
not always the case. In Fig.~\ref{fig:logBy} we compare spatiotemporal evolutions of $B_{y}(z,t)$, in a
longer box ($L_{z}=32$) with less driving noise, which causes a lower growth 
rate and a decrease in the relative important of the stochastic-$\alpha$ effect compared to the coherent dynamo.  
As is evident, the two dynamos are qualitatively 
different, with the phase of $B_{y}$ wandering quasi-randomly in the non-rotating case, while
in the rotating case it is approximately constant in time. This constant
phase is not consistent with a dynamo driven purely by the stochastic $\alpha$ effect (see Sec.~\ref{sec:Shear-dynamos}).

From Figs.~\ref{fig:Rotating low Rm} and \ref{fig:logBy}, we thus interpret the 
Keplerian rotating shear dynamo around these parameters as being driven by both
incoherent and coherent mechanisms. This interpretation is
entirely consistent with all numerical results given in \citet{Yousef:2008ix,Yousef:2008ie}.
In particular, their Fig.~5 illustrates that the addition of rotation
enhances the growth of the dynamo in all cases. Furthermore, while
$\gamma\sim S$ for the non-rotating dynamo, with rotation it is evident
that the growth of $\gamma$ is somewhat faster than linear in $S$. Since
one expects $\gamma\sim S^{2}$ for a coherent dynamo (since $\eta_{yx}$
itself must scale linearly with $S$ for small $S$), their observed
trends are consistent with the dynamo being driven through a mix of incoherent and
coherent mechanisms. We note that a consideration of wider
boxes would increase the importance of the coherent effect in comparison to the incoherent
effect, widening the difference between rotating and non-rotating dynamos.

\paragraph*{Varying the rotation}
As one final test of the importance of net rotation in this system
we have run a series of simulations, increasing the rotation from
$\Omega=-1$ (cyclonic rotation) to $\Omega=4$ (anticyclonic rotation).
\begin{figure}
\begin{centering}
\includegraphics[width=0.8\columnwidth]{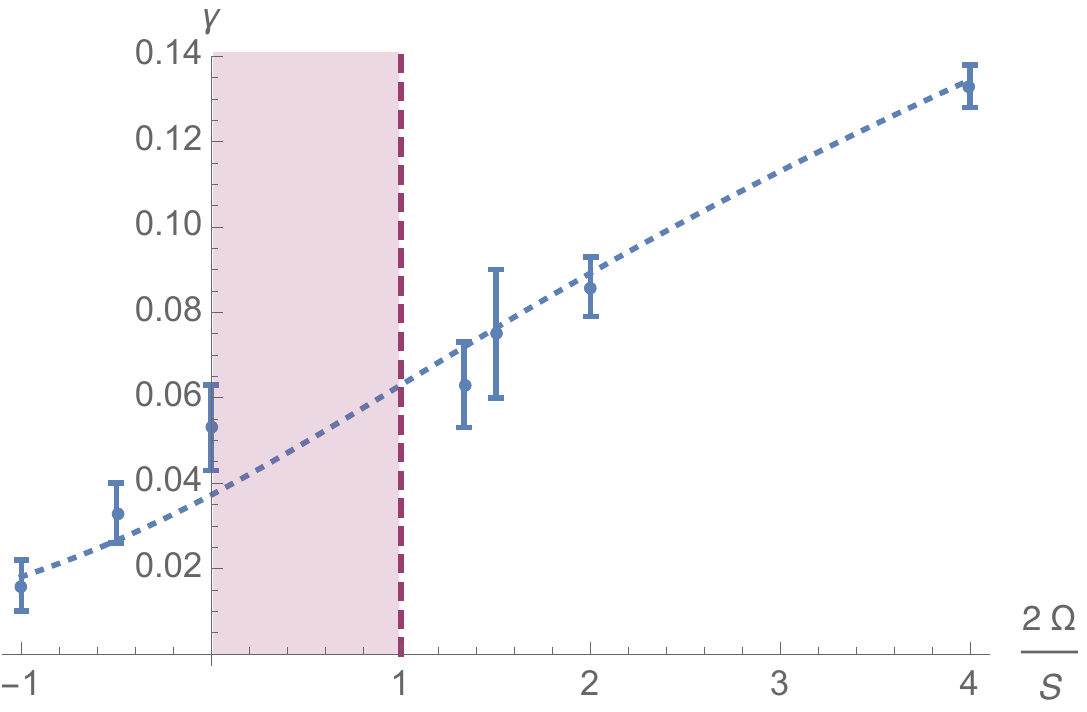}

\caption{Growth rate of $B_{rms}$ as a function of $2\Omega/S$, for fixed
shear $S=2$, and velocity driving $u_{rms}^{2}\approx1$ (except
for the point $\Omega=0$, for which $u_{rms}^{2}\approx1.5$). The
shaded region shows where the flow is hydrodynamically unstable (neglecting
dissipation), and the dashed vertical line shows the SOCA prediction
for where the coherent dynamo growth rate vanishes. Of course, due
to the strong stochastic alpha effect, the dynamo can still grow even
when the predicted coherent growth rate is zero or negative. The dotted
line is an approximate fit of predicted growth rate, Eq.~\eqref{eq:gamma sa and sc}, to
 the data, using $\eta_{t}=u_{rms}/3k_{f}=0.018$, $\eta_{yx} =0.0007\times(2\Omega-S)$, and
 $\langle \alpha_{yy}^{2}\rangle=6.2\times 10^{-5}$. Error bars are
estimated by fitting the growth rate to half of the time-series data for each run. \label{fig:changing rotation}}
\end{centering}

\end{figure}
Results from this series of simulations are illustrated in Fig.~\ref{fig:changing rotation}.
As expected, we see a substantial increase in dynamo growth rate as
the rotation becomes anticyclonic, in broad agreement with the SOCA prediction
$\eta_{yx}\propto S-2\Omega$. Due to the  presence of the stochastic
$\alpha$ effect, one would not expect a linear scaling of $\gamma$ with $\Omega$. Instead,
the growth rate (including an $\eta_{yx}$ and fluctuating $\alpha_{yy}$)
is the most positive root of\footnote{In reality, one will also see a change in growth rate due to $\eta_{xy}$ and fluctuations in the other $\alpha$ coefficients, but these effects seem  minor and are ignored here.}
\begin{equation}
-4\left\langle \alpha_{yy}^{2}\right\rangle k^{2}S^{2}-4\eta_{yx}k^{2}S\xi+\xi^{3}=0,\label{eq:gamma sa and sc}
\end{equation}
where $\xi=2\eta_{t}k^{2}+\gamma$ \citep{Mitra:2012iv}.
We plot a fit of Eq.~\eqref{eq:gamma sa and sc} to the data in Fig.~\ref{fig:changing rotation} [with 
$\eta_{t}=u_{rms}/3k_{f}$, $\eta_{yx} = \eta_{yx0}(2\Omega-S)$], illustrating good agreement away from the instability boundaries 
($\Omega=0$ and $\Omega = S/2$). Close to the boundary,  it seems that some 
other nonlinear effect may be important, increasing the growth rate on the
cyclonic side and decreasing it on the anticyclonic side.

\subsection{Direct calculation of transport coefficients\label{sub:Direct-calculation-of}}

To validate and quantify the conclusions discussed above, in this section we directly
calculate the transport coefficients, comparing results from CE2\footnote{DQLS gives identical 
results to CE2, albeit with errors due to the random noise.} and the test-field method \citep{Brandenburg:2005kla} 
(implemented within the framework of the SNOOPY code). 
The CE2 calculations are carried out by fixing the mean fields, $B_{y}=B_{y0} \cos(2\pi z/L_{z})$,
and driving linear fluctuations to calculate their statistics and thus the transport coefficients. 
These calculations, since they are quasi-linear
in the shearing frame, are fundamentally the same as those presented 
in \citet{Sridhar:2009jg,Sridhar:2010it,SINGH:2011ho} for the non-rotating case.\footnote{The only substantial
difference is the forcing -- a singular forcing $\bm{\sigma}_{\bm{u}}\sim\delta\left(\bm{k}-\bm{k}_{f}\right)$
is used in \citet{SINGH:2011ho}, while we use the same forcing as
detailed in Sec.~\ref{sec:Equations etc.}. Of course, our calculation is numerical rather than
analytic and it is trivial to add the effects of rotation (this
is difficult analytically although perturbative methods may be feasible,
see \citealt{Leprovost:2008jm}).}

Test-field method calculations are carried out in the standard way  \citep{Brandenburg:2008bc} by solving the momentum
equation with no Lorentz force,  using this velocity field to drive a small-scale magnetic induction equation
\begin{equation}
\partial_{t}\bm{b}^{q} = \nabla\times \left[\bm{u}\times\bm{B}^{q}+\bm{U}\times\bm{b}^{q}+(\bm{u}\times\bm{b}^{q}-\overline{\bm{u}\times\bm{b}^{q}})\right]+\bar{\eta}\triangle\bm{b}
\end{equation}
for a set of test-fields $\bm{b}^{q}$ and specified mean-fields $\bm{B}^{q}$ (we chose a sinusoidal form for $\bm{B}^{q}$). 
There is no small-scale dynamo at
these parameters, which simplifies the calculation since $\bm{b}^{q}$ arises purely due to the presence of $\bm{B}^{q}$.
Calculations are run from $t=0\rightarrow1000$, with the error in the transport coefficients estimated by 
dividing the time-series into $N$ segments and calculating the standard deviation of the mean.\footnote{We generally take $N=100$, but
results are quite insensitive to this choice.}

\begin{figure}
\begin{centering}
\includegraphics[width=1.0\columnwidth]{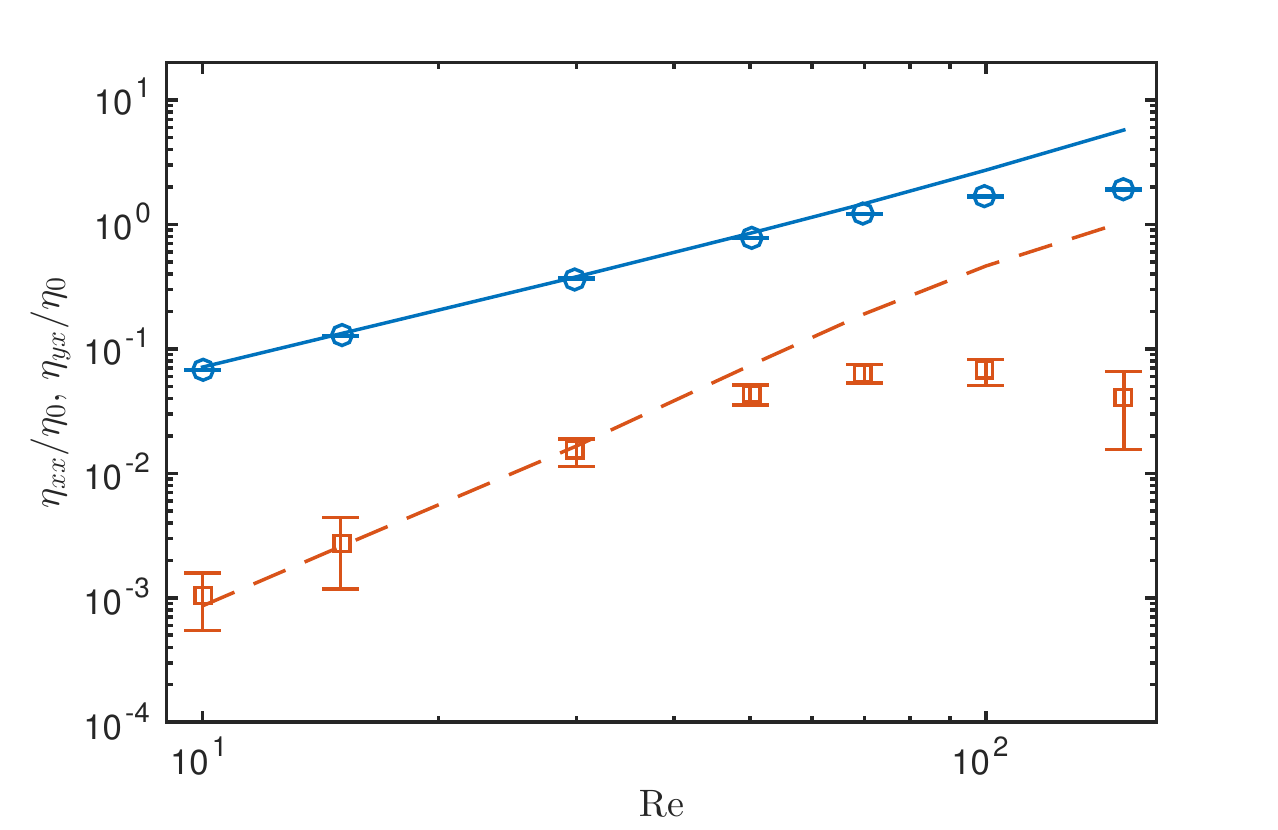}

\caption{Transport coefficients for the kinematic non-rotating dynamo $\eta_{xx}$ (solid line and circle markers,
blue) and $\eta_{yx}$ (dashed line and square markers, black) as a function of Re (at Pm=1), for $S=2$. 
The curves show the quasi-linear results, calculated using CE2, while the markers show the 
nonlinear test-field calculations with error bars (see text).
As is common, coefficients are normalized by the ``high-conductivity'' SOCA turbulent resistivity $\eta_{0}=u_{rms}/(3k_{f})$.
Across all simulations, the absolute level of the forcing (i.e., $\bm{\sigma}_{\bm{u}}$) is kept constant at the same level as Fig.~\ref{fig:Non rotating low Rm}, which means that the lower $\mathrm{Re}$ simulations have somewhat lower $u_{rms}$. \label{fig:kinematic transport Nrot} }
\end{centering}
\end{figure}

\begin{figure}
\begin{centering}
\includegraphics[width=1.0\columnwidth]{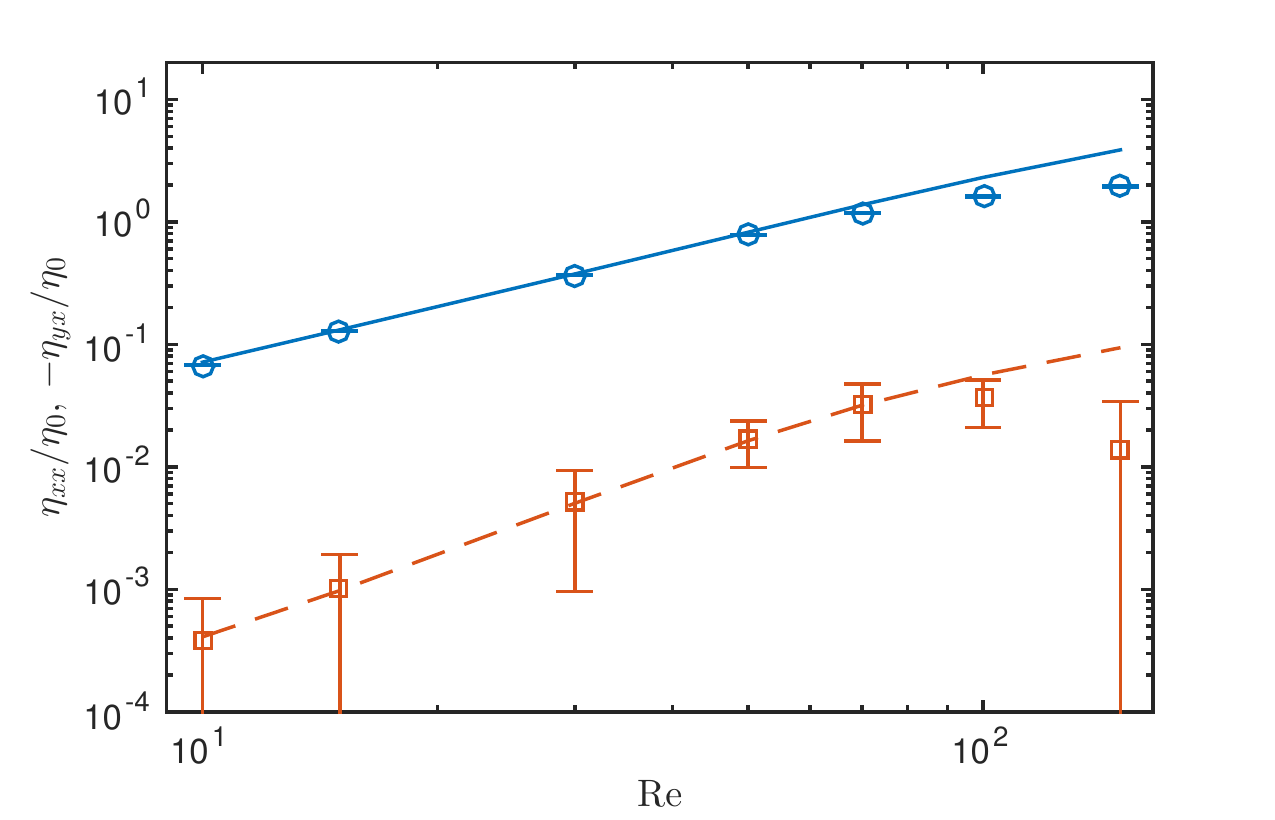}

\caption{Same as for Fig.~\ref{fig:kinematic transport Rot} but with Keplerian rotation. Note
that $\eta_{yx}<0$ in this case, so we plot $-\eta_{yx}$ so as to utilize a $\log$ scale. \label{fig:kinematic transport Rot}}
\end{centering}
\end{figure}

Results are illustrated in Figs.~\ref{fig:kinematic transport Nrot} and \ref{fig:kinematic transport Rot}.
We see that in both cases the quasi-linear and nonlinear coefficients agree
at lower $\mathrm{Rm}$ as expected, diverging somewhat past $\mathrm{Rm}\gtrsim 70$.
In agreement with our conclusions from simulations earlier in the section,  
$\eta_{yx}>0$ in the non-rotating case, while $\eta_{yx}<0$ with rotation, showing that a
coherent dynamo is possible at sufficiently small $k_{z}$. It is also worth noting
that the magnitude of $\eta_{yx}$ is less in the rotating case, as known from SOCA calculations \citep{Radler:2006hy,Analytic}.
For $\mathrm{Rm}=100$, as used in Figs.~\ref{fig:Non rotating low Rm} to \ref{fig:changing rotation}, there 
are some differences between quasi-linear and nonlinear results due to inaccuracies in the 
quasi-linear approximation, which explains the discrepancy in dynamo growth rates observed\footnote{
It seems that for Fig.~\ref{fig:Non rotating low Rm}, the lower values of $\eta_{xx}$ and $\eta_{yx}$ in comparison
to the quasi-linear runs cancel each other, leading to the same growth rate.} in Fig.~\ref{fig:Rotating low Rm}.
Interestingly, given the controversies surrounding 
the kinematic ``shear-current'' effect \citep{Rogachevskii:2003gg}, nonlinear 
corrections appear to be particularly important for $\eta_{yx}$ without rotation (this coefficient shows the largest 
discrepancy between the nonlinear and quasi-linear calculations).

In addition to the results for $\eta_{yx}$ and $\eta_{xx}$ shown,
we have also calculated $\eta_{xy}$ and $\eta_{yy}$ by setting $B_{x}=B_{x0}\cos\left(k_{1}z\right)$,
$B_{y}=0$. We find that $\eta_{xx}=\eta_{yy}$ to a high degree
of accuracy, while $\eta_{xy}$, which is positive, is mostly unaffected by rotation.
Its magnitude (compared to the other $\eta$) depends strongly
on the shear and Reynolds number. Due to the dominance of the shear,
such an $\eta_{xy}$ has little effect on the growth rate, even though
its magnitude is larger than that of $\eta_{yx}$. In addition to the results illustrated and discussed above, we
have also verified the expected linear dependence
of $\eta_{yx}$ on $S$ at low $\mathrm{Rm}$ and confirmed that the 
transport coefficients change very little with $L_{z}$ over the range $L_{z}=1\rightarrow8$.

\section{Magnetically driven dynamo}\label{Magnetic}

Having now broadly understood the shear dynamos of \citet{Yousef:2008ie,Yousef:2008ix},
we examine the effect of small scale magnetic fluctuations. Before
presenting numerical results, it is helpful to explain in more detail exactly
what we is meant by a magnetically driven linear dynamo. Similar ideas
have been considered before (e.g., \citealt{Radler:2003gg,Park:2012eg}), see 
\citet{Rheinhardt:2010do} for a particularly thorough analysis.

\begin{figure}
\includegraphics[width=0.965\columnwidth]{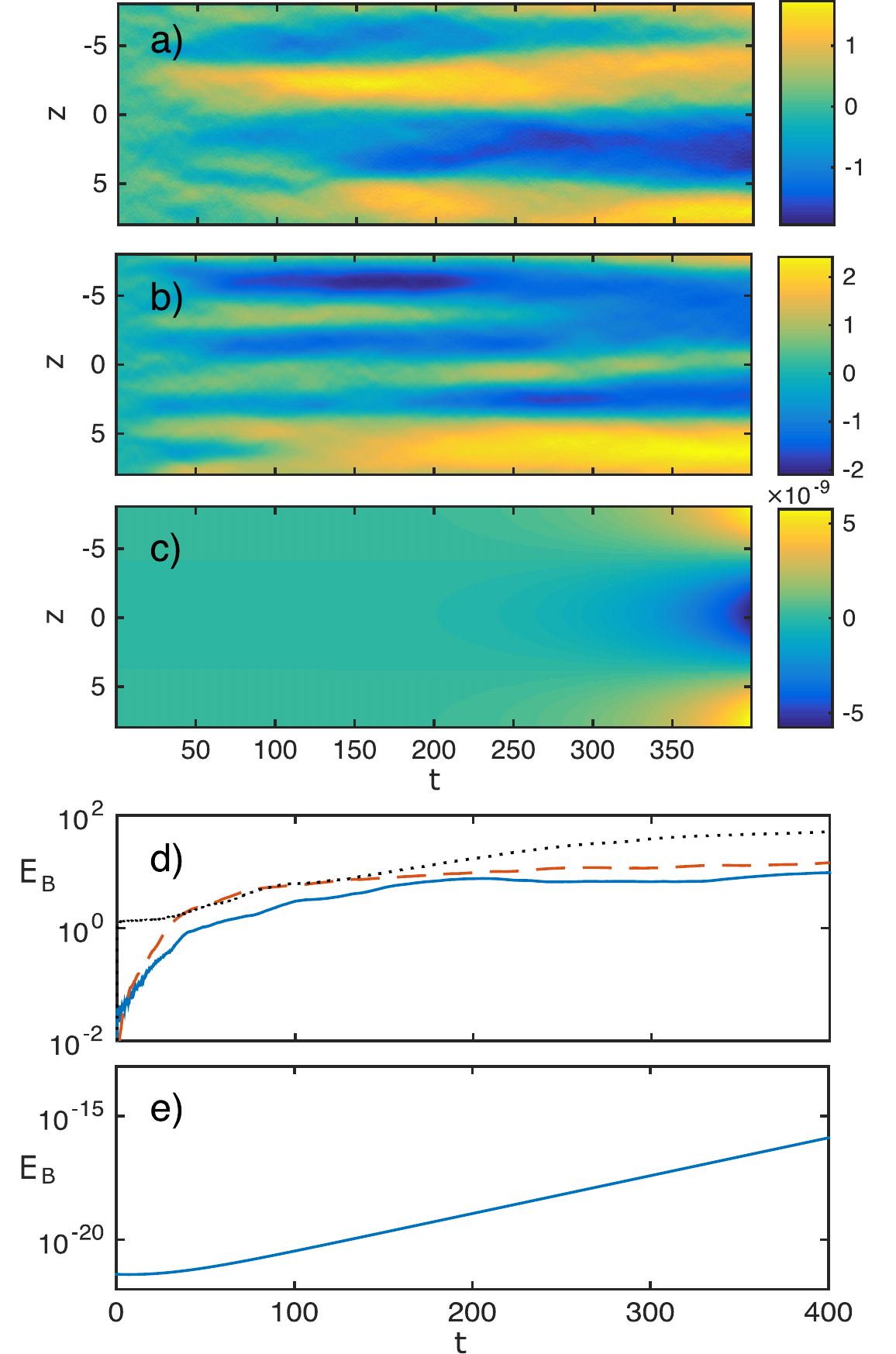}
\caption{(a-c) Illustration of $B_{y}\left(z,t\right)$ from non-rotating turbulence, using 
$\bm{\sigma}_{\bm{b}}=\bm{\sigma}_{\bm{u}}$
with $S=2,$ $L_{z}=16$, for DNS, DQLS, and CE2 from (a)-(c). 
These parameters are identical to Fig.~\ref{fig:Non rotating low Rm}, but
with $\bm{\sigma}_{\bm{b}}=\bm{\sigma}_{\bm{u}} = \bm{\sigma}/2$, where $\bm{\sigma}$ is 
the driving noise used in Fig.~\ref{fig:Non rotating low Rm}.
Part (d) shows the growth in time of
the mean-field for (solid,blue) nonlinear equations;  (dashed,red) quasi-linear
DNS (the dotted black line shows the fluid energy in the nonlinear run). (e) mean-field growth from CE2. 
As shown by the CE2 growth, the coherent dynamo is much stronger than in the kinematic case, but
because the direct simulations (a-b) start from high amplitudes, it is hard to see
the exponential dynamo growth phase in these simulations. 
\label{fig:mag comparison NRot}}
\end{figure}

As is obvious from Eq.~\eqref{eq:genMF}, an unstable dynamo
requires inhomogeneity in the fluctuations $\bm{u}$ and $\bm{b}$,
such that $\nabla\times\mathcal{E}\neq0$. Since we assume initially
homogenous fluctuations (termed $\bm{u}_{0}$ and $\bm{b}_{0}$),
this inhomogeneity must be introduced by $\bm{B}$, which is assumed
small. Considering the linearized fluctuation equations for simplicity
(this is just the quasi-linear dynamo, which we know works in any
case),\footnote{Inclusion of
nonlinear terms introduces several additional complexities, see \citet{Rheinhardt:2010do}.} 
it is evident that the kinematic dynamo arises from inhomogeneity
induced in $\bm{b}$ fluctuations through the term $\nabla\times\left(\bm{u}_{0}\times\bm{B}\right)$
in the fluctuation induction equation {[}Eq.~\eqref{eq:QL fluct eqs b}{]}.
This leads to an inhomogenous contribution to $\mathcal{E}$ through
$\bm{u}_{0}\times\bm{b}_{\mathrm{inhom}}$. In contrast, in the presence
of $\bm{b}_{0}$, an inhomogenous part of $\bm{u}$ will arise from
the Lorentz force $\bm{b}_{0}\cdot\nabla\bm{B}+\bm{B}\cdot\nabla\bm{b}_{0}$
{[}see Eq.~\eqref{eq:QL fluct eqs u}{]}, giving a contribution to
$\mathcal{E}$ through $\bm{u}_{\mathrm{inhom}}\times\bm{b}_{0}$.
Without a mean-field flow, such a contribution is not possible from
the induction equation alone. In calculating the transport coefficients
(Secs.~\ref{sub:Direct-calculation-of} and \ref{sub:Direct-calculation-magnetic})
we have verified that artificial removal of the Lorentz force causes
the transport coefficients to return to their kinematic values.
It may be interesting in future work to examine in the vorticity dynamo
(i.e., generation of $\bm{U}$) in more detail, in particular its
interaction with the magnetic dynamo \citep{Courvoisier:2010bj}. These effects are almost certainly
much more important in the non-rotating case.

Before proceeding it is worth commenting on an important difference between the magnetic shear-current
effect discussed below and the standard magnetic $\alpha$-effect. 
This difference stems from the fact that the magnetic $\alpha$-effect can have either sign, since it is related to the 
small-scale current helicity, $\alpha_{M}\sim -\left<\bm{b}\cdot \nabla \times \bm{b}\right>$. In practice, 
as the small-scale dynamo grows in the presence of helical velocity fluctuations, $\alpha_{M}$ grows 
with the \emph{opposite sign} to the kinematic $\alpha$-effect\footnote{
One possible exception to this that may be very important could occur in the presence
of magnetic instabilities, for instance the MRI. In this case it seems more likely that the magnetic
$\alpha$ effect might overwhelm the kinematic effect, since $\bm{b}$ fluctuations do not arise purely as a consequence
of small-scale dynamo action \citep{Gressel:2010dj,Park:2012eg}.}
 -- the origin of catastrophic quenching \citep{Blackman:2002fe,Brandenburg:2005kla}.
In contrast, since the magnetic shear-current effect drives the dynamo through a resistivity, $\eta\sim \langle\bm{b}^{2}\rangle$, its sign is fixed. 
This implies that the the source of magnetic fluctuations can be the small-scale dynamo, in some sense
the inverse of quenching. In a recent
paper \citep{HighRm}, we have shown that this mechanism is realizable at higher
$\mathrm{Rm}$ where the small-scale dynamo is unstable. In particular, we see
a  decrease in $\eta_{yx}$ after saturation of the small-scale dynamo, which can
in turn  drive a coherent large-scale dynamo.

\begin{figure}
\includegraphics[width=\columnwidth]{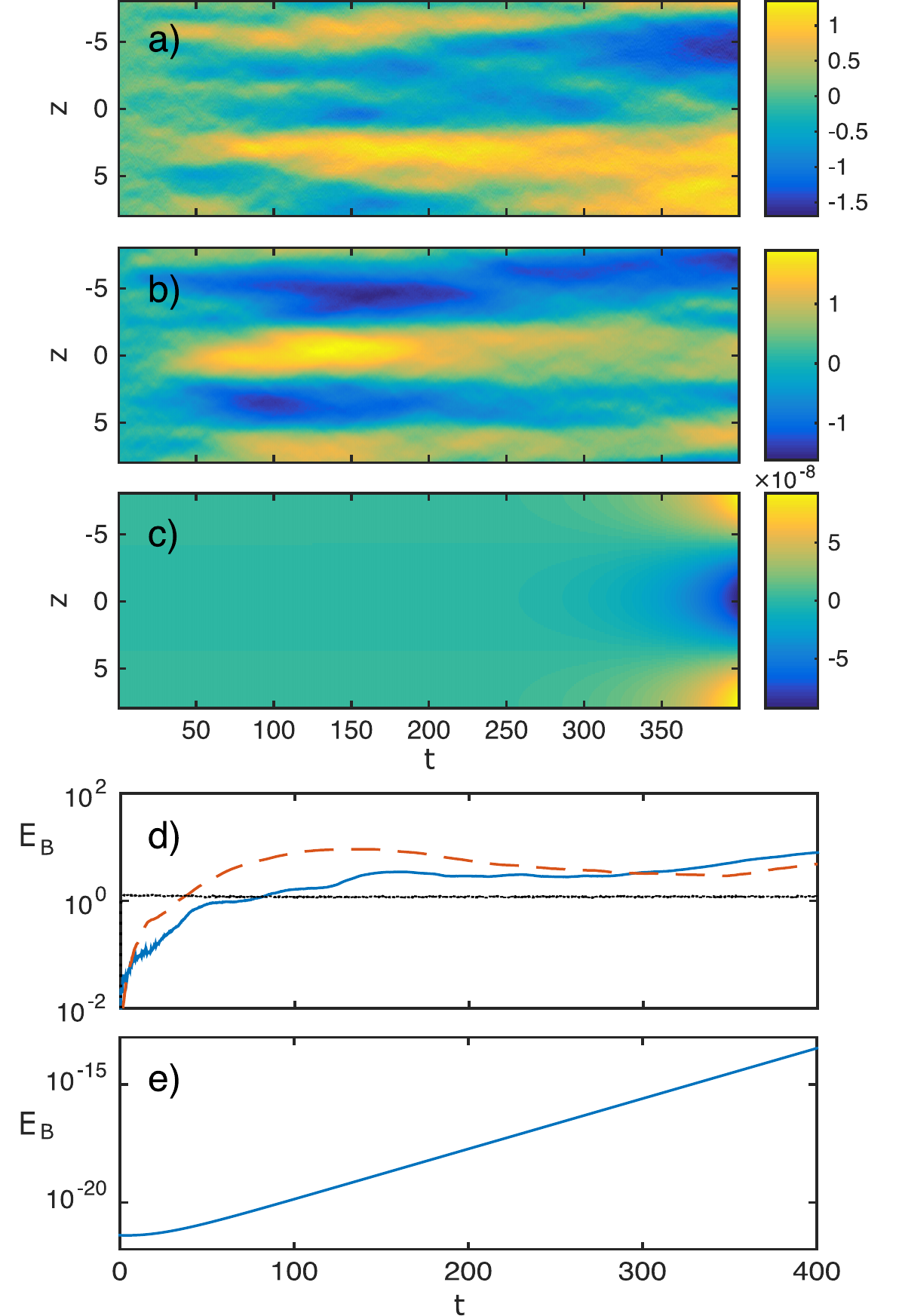}
\caption{Same as Fig.~\ref{fig:mag comparison wRot}, but using Keplerian
rotation (cf. Fig.~\ref{fig:Rotating low Rm}). The behavior is broadly similar to 
Fig.~\ref{fig:mag comparison wRot}, with a slightly higher 
coherent growth rate because the velocity fluctuations have a positive
effect in this case. \label{fig:mag comparison wRot}}

\end{figure}

\subsection{Numerical experiments on the magnetic dynamo}
Here, we argue for the existence of the magnetic dynamo through numerical experiments, in a similar 
way to the discussion in Sec.~\ref{Kinematic}.
Since there is no small-scale dynamo due to the low Reynolds numbers, we excite homogenous magnetic fluctuations 
($\bm{b}_{0}$) by forcing the induction equation with $\bm{\sigma}_{\bm{b}}$ (the
statistical properties of which are chosen to be the same as $\bm{\sigma}_{\bm{u}}$).

Figures~\ref{fig:mag comparison NRot} and \ref{fig:mag comparison wRot} illustrate
the comparison between DNS, DQLS, and CE2, at identical parameters to Figs.~\ref{fig:Non rotating low Rm} and \ref{fig:Rotating low Rm} and using the same total forcing level (i.e., $\bm{\sigma}_{\bm{b}}=\bm{\sigma}_{\bm{u}} = \bm{\sigma}/2$, where $\bm{\sigma}$ is 
the driving noise used in Figs.~\ref{fig:Non rotating low Rm} and \ref{fig:Rotating low Rm}).
The most obvious difference -- comparing Figs.~\ref{fig:mag comparison NRot} and \ref{fig:mag comparison wRot} to Figs.~\ref{fig:Non rotating low Rm} and \ref{fig:Rotating low Rm} --  is the much higher
amplitudes in the direct numerical simulations (both quasi-linear and nonlinear). 
This is not due to the mean-field dynamo and simply results from the 
approximate equipartition of $\bm{B}$ with $\bm{b}$ due to the finite size of the domain. This 
occurs almost immediately because of the strong $\bm{b}_{0}$. Thus, the strong magnetic
fields observed at later times in the direct simulations are in the nonlinear
saturation regime of the dynamo, where $\eta_{yx}$ might be expected to change sign.
Since this paper is concerned with the linear growth phase, we shall
not analyze this  saturation phase in detail. 

In contrast, CE2, by eliminating all fluctuations in $\mathcal{E}$, allows the mean-field exponential growth 
phase to be observed despite the presence of strong magnetic fluctuations. 
(We remind the reader here that the fundamental model used in CE2 is identical to
 DQLS, the only difference arises from statistics being directly inserted into $\mathcal{E}$ to drive the
 mean-fields.) Comparing with Figs.~\ref{fig:Non rotating low Rm}(d) and \ref{fig:Rotating low Rm}(d), we see that 
 in both cases the mean-field growth is substantially faster; that is, the magnetic fluctuations are contributing
 significantly to mean-field growth through the shear-current effect. In fact,
 since the growth rate is still strong in the non-rotating case [Fig.~\ref{fig:mag comparison NRot}(e)], 
it is clear that the magnetic $\eta_{yx}$ 
significantly overwhelms the (positive) kinematic
 $\eta_{yx}$, for the same forcing level $\bm{\sigma}_{\bm{u}}=\bm{\sigma}_{\bm{b}}$. 
 Thus we have a mean-field dynamo driven by the magnetic shear-current effect. Moreover, the
magnetic fluctuations produce a stronger dynamo driving than the velocity fluctuations.
 
 Quantitative comparison of the CE2 calculations with the direct simulations is not  possible
 due to the importance of nonlinear effects on the dynamo in the direct runs. While 
 there  appears to be an exponential growth phase in each direct case (before $t\approx 100$),
 it is somewhat too short to say for sure. In both the nonlinear and quasi-linear
 runs the wavelength appears to be significantly shorter ($k_{z}\approx 4\pi/L_{z} \rightarrow 6\pi/L_{z}$) than in CE2, for which the growing mean-field is the largest mode in the box. 
  Since a stochastic-$\alpha$ effect is expected to be important (presumably at a similar
 level to the kinematic case), this is not surprising; $\alpha$ fluctuations will act to increase the growth 
 rate, decreasing the wavelength of the most unstable mean-field mode. The
 nonlinear and quasi-linear evolutions are broadly similar, although the nonlinear
 runs saturate at slightly lower amplitudes than the quasi-linear
 cases (a detailed comparison is 
 not possible without running an ensemble of such simulations). 
 Given this similarity -- combined with the knowledge that the DQLS mean-field
 is, at least partially, driven by a coherent effect (the CE2 dynamo is unstable) -- 
 we conclude that this magnetic shear-current effect should also be playing a significant
 role in both the rotating and nonrotating direct numerical simulations [Figs.~\ref{fig:mag comparison NRot}(a) and \ref{fig:mag comparison wRot}(a)].

\paragraph*{Nonlinear DNS}
\begin{figure*}
\centering
\includegraphics[width=\textwidth]{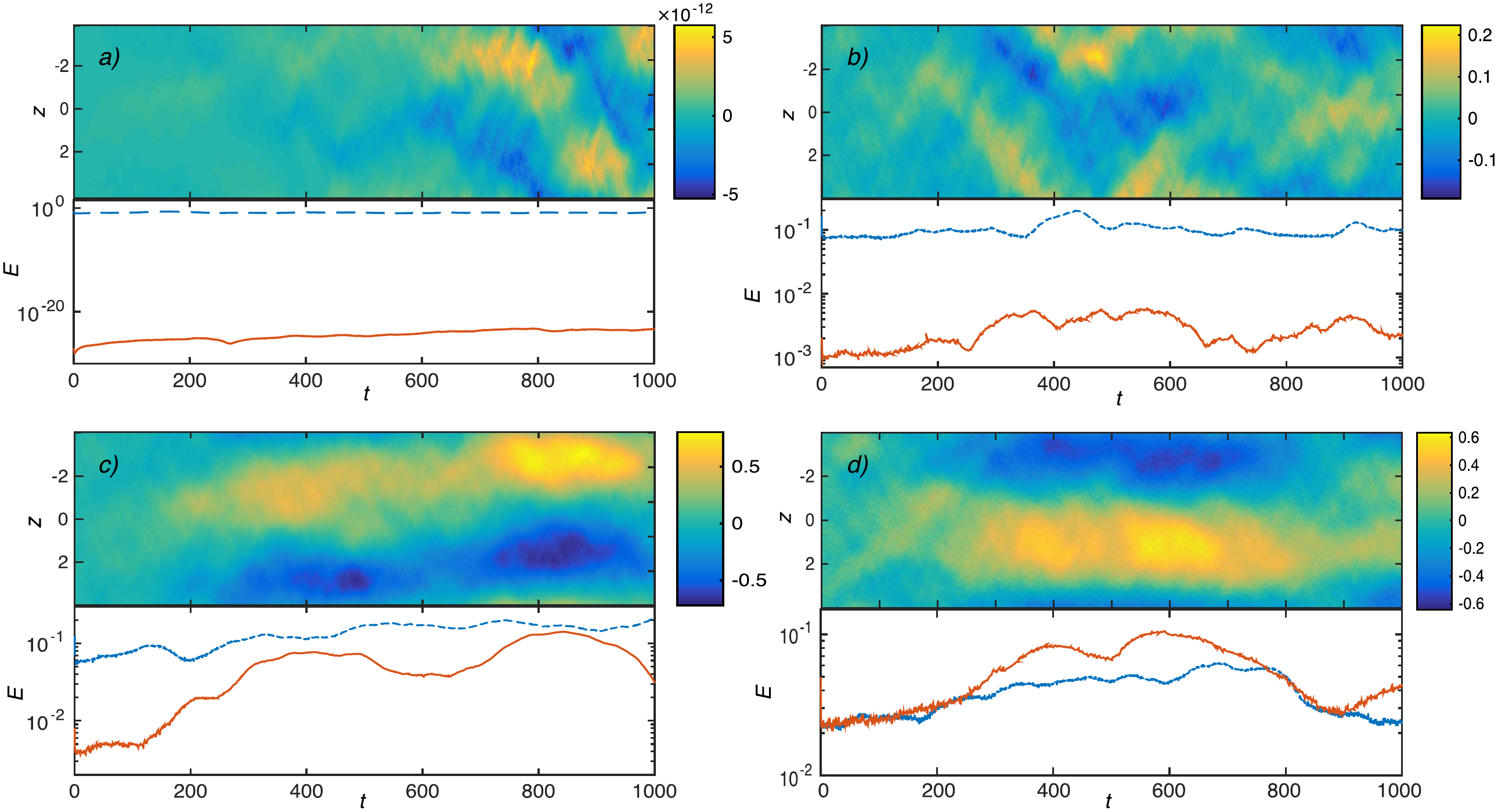}
\caption{Low Rm driven DNS at $S=1$, $L_{z}=8,$ and no rotation, with $\bm{\sigma}_{\bm{u}}+\bm{\sigma}_{\bm{b}}=\bm{\sigma}$
chosen to be constant in each simulation (the level is such that $u_{rms}\approx0.2$
when only velocity forcing is used). From left to right we take (a) $\bm{\sigma}_{\bm{b}}=0$,
(b) $\bm{\sigma}_{\bm{b}}=0.1\bm{\sigma}$, (c) $\bm{\sigma}_{\bm{b}}=0.2\bm{\sigma}$
(d) $\bm{\sigma}_{\bm{b}}=0.5\bm{\sigma}$. The top row of each subfigure
illustrates the time development of $B_{y}\left(z,t\right)$, the
bottom row illustrates the kinetic energy (dashed, blue) and magnetic
energy (solid, red).\label{fig:Low-Rm-magnetic driven}}

\end{figure*}

Knowing that the magnetic fluctuations can drive the coherent mean-field
dynamo, it is helpful to examine the qualitative changes that occur as $\bm{b}_{0}$ is increased.
As a simple numerical experiment with the nonlinear equations, we start from pure velocity forcing
and increase the driving in the induction equation, while
keeping the total forcing, $\bm{\sigma}_{\bm{u}}+\bm{\sigma}_{\bm{b}}=\bm{\sigma}$,
fixed. (While we have carried out these experiments both with and without Keplerian rotation, we 
present only the non-rotating cases here as the rotating results are similar.)  

Results in the range $\bm{\sigma}_{\bm{b}}=0\rightarrow0.5\bm{\sigma}$
are illustrated in Fig.~\ref{fig:Low-Rm-magnetic driven}. At $\bm{\sigma}_{\bm{b}}=0$,
we see a similar dynamo to that in Fig.~\ref{fig:Non rotating low Rm}, although it is a little weaker due to the lower $u_{rms}$
and choice $S=1$.
This is a stochastic-$\alpha$ effect, as seen by the slowly growing
mean-fields that wander significantly in phase. 
Let us now consider the more interesting behavior of the other cases, $\bm{\sigma}_{\bm{b}}=0.1\bm{\sigma}$,
$\bm{\sigma}_{\bm{b}}=0.2\bm{\sigma}$ and $\bm{\sigma}_{\bm{b}}=0.5\bm{\sigma}$. Firstly,  note that
the larger mean-fields compared to the kinematic case are purely due to 
equipartition of $\bm{B}$ with $\bm{b}$, as in Fig.~\ref{fig:mag comparison NRot}(a). Instead, our main result is the substantial
qualitative difference in the appearance of the mean-field evolution between $\bm{\sigma}_{\bm{b}}=0.1\bm{\sigma}$
and the cases with higher magnetic forcing. Specifically,  at $\bm{\sigma}_{\bm{b}}=0.1\bm{\sigma}$
one observes a wandering mean-field as well as possibly a slow growth, behavior we interpret 
as a stochastic-$\alpha$ effect near its saturated
state. In contrast, at $\bm{\sigma}_{\bm{b}}=0.2\bm{\sigma}$ and $\bm{\sigma}_{\bm{b}}=0.5\bm{\sigma}$,
a relatively fast growth of $\bm{B}$ is observed until
saturation at substantially larger values than seen at $\bm{\sigma}_{\bm{b}}=0.1\bm{\sigma}$.
In addition, the profile of $B_{y}\left(z,t\right)$ for  $\bm{\sigma}_{\bm{b}}\gtrsim0.2\bm{\sigma}$ is relatively
constant in phase, suggesting that the dynamo is coherent.

This behavior again suggests that a coherent dynamo can be driven by small-scale
magnetic fluctuations -- the magnetic shear-current effect. This
dynamo saturates at larger field strengths than the stochastic-$\alpha$
dynamo, with the saturation amplitude being roughly independent of
the level of magnetic fluctuations {[}as seen by comparison\footnote{Note that the
different $z$ domain size must be taken into account to compare Fig.~\ref{fig:Low-Rm-magnetic driven} with Fig.~\ref{fig:mag comparison NRot}(a).}
 of Figs.~\ref{fig:Low-Rm-magnetic driven}(c)
and (d){]}. Note also that this dynamo field appears to show quasi-cyclic
behavior of some sort in its nonlinear regime [in Fig.~\ref{fig:Low-Rm-magnetic driven}(d)
the large scale field reappears again at later times]. The reason for 
this interesting behavior and its relevance to other dynamo cycles (e.g., in
MRI turbulence, \citealt{Lesur:2008cv}) remains unclear, and given
its apparent origin in nonlinear dynamo physics, we leave its study
to future work. 

To ensure the observed behavior is robust, we have rerun each of the simulations in
Fig.~\ref{fig:Low-Rm-magnetic driven} several times, varying the initial
conditions and random number seed. These (not shown) have illustrated that the $\bm{\sigma}_{\bm{b}}=0.1\bm{\sigma}$
simulations occasionally excite the coherent dynamo similar
to that in Fig.~\ref{fig:Low-Rm-magnetic driven}(c-d), and will eventually do so if 
evolved for a sufficiently long time. In addition, the $\bm{\sigma}_{\bm{b}}=0.2\bm{\sigma}$
occasionally \emph{fails} to excite the coherent dynamo as quickly as observed in Fig.~\ref{fig:Low-Rm-magnetic driven}(c). 
This brings us to the conclusion that the coherent dynamo can be excited
for $\bm{\sigma}_{\bm{b}}\gtrsim(0.1\rightarrow 0.2)\times \bm{\sigma}$ and
the simulation outcome depends on  properties of an individual realization around this boundary. 
We have failed to find coherent dynamo excitation
at $\bm{\sigma}_{\bm{b}}=0.05\bm{\sigma}$, having tested a number
of realizations over very long time periods. 
This dependence on realization is very similar to the behavior observed in
shear-dynamos at higher $\mathrm{Rm}$, where the small-scale 
dynamo acts as the source of $\bm{b_{0}}$ fluctuations \citep{HighRm}.

\subsection{Direct calculation of transport coefficients\label{sub:Direct-calculation-magnetic}}

As in Sec.~\ref{sub:Direct-calculation-of} we can directly calculate
the transport coefficients of the magnetic dynamo by fixing the
mean-fields and driving magnetic fluctuations. Within
quasi-linear theory, this is a straightforward generalization of
kinematic calculations, and the transport coefficients
in the presence of both magnetic and velocity fluctuations will be the
sum of those calculated with one or the other, $\eta=\eta_{u}+\eta_{b}$.  However, 
inclusion of magnetic fluctuations
in the nonlinear test-field method can be more complex \citep{Rheinhardt:2010do}, and 
the linearity of the transport coefficients is lost, $\eta\neq\eta_{u}+\eta_{b}$ (although of course
at  low Rm nonlinear results must approach the quasi-linear results). Because
of this, we present only quasi-linear results for the magnetic dynamo coefficients, and leave
magnetic test-field method studies to future work.
\begin{figure}
\begin{centering}
\includegraphics[width=1.0\columnwidth]{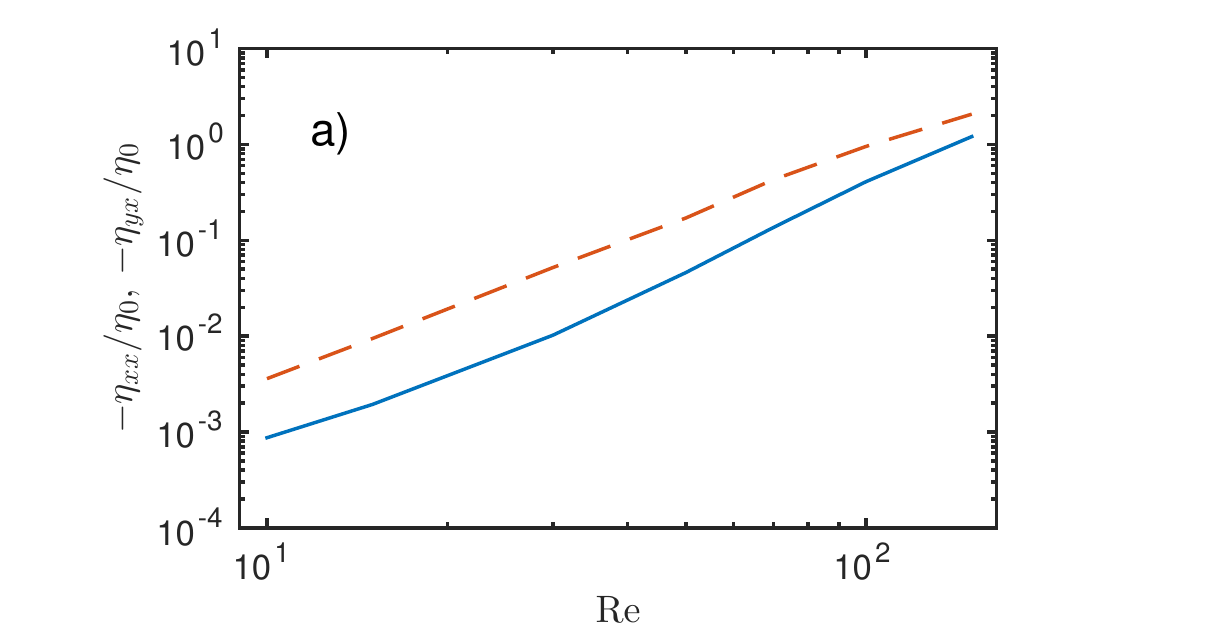}
\includegraphics[width=1.0\columnwidth]{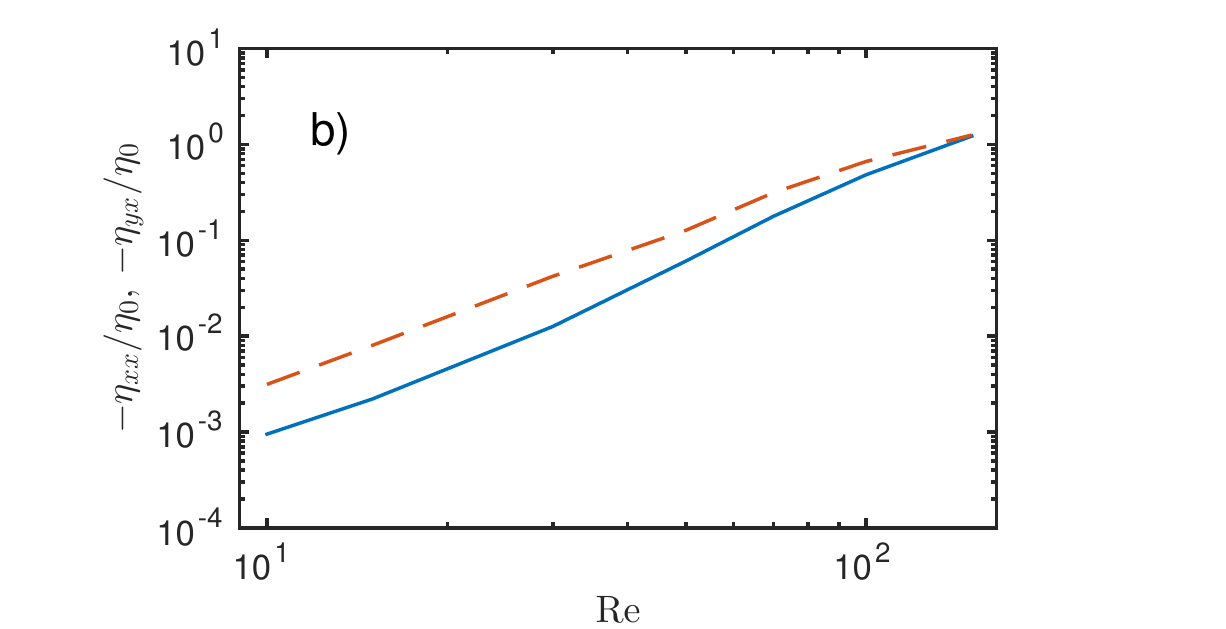}

\caption{Transport coefficients for the magnetic dynamo $-\eta_{xx}$ (solid,
blue) and $-\eta_{yx}$ (dashed, red)  as a function of $\mathrm{Re}$ (at $\mathrm{Pm}=1$), for $S=2$ and (a) $\Omega=0$
(b) Keplerian rotation $\Omega=4/3$. (Note that both $\eta_{xx}$ and $\eta_{yx}$ 
are negative). The calculations are carried
out at $L_{z}=4$ using CE2 (as for the kinematic case, there 
is very little dependence of $L_{z}$). Coefficients are normalized by 
the $u_{rms}$ values from Figs.~\ref{fig:kinematic transport Nrot} and \ref{fig:kinematic transport Rot}, such 
that the values of the $u$ and $b$ transport coefficients can be directly compared. 
\label{fig:b Transport-coefficients}}
\end{centering}

\end{figure}

Fig.~\ref{fig:b Transport-coefficients} illustrates $\eta_{yx}$ and
$\eta_{xx}$ when only magnetic fluctuations are present, as calculated
by setting $\bm{\sigma}_{\bm{u}}=0$ and fixing $B_{y}$ in CE2, with the
same technique as detailed in Sec.~\ref{sub:Direct-calculation-of}.
Most notably, we see both that $\eta_{yx}$ and $\eta_{xx}$ are negative, both
with and without rotation,
and are of similar magnitudes.  Importantly, a comparison of Fig.~\ref{fig:b Transport-coefficients} with 
Figs.~\ref{fig:kinematic transport Nrot}-\ref{fig:kinematic transport Rot} shows 
that $\eta_{yx}$ is substantially larger in magnitude than the kinematic value,
which implies that when $b_{rms}\sim u_{rms}$ the magnetic contribution 
should dominate. For example, without rotation, the quasi-linear magnetically driven $\eta_{yx}$
is larger than the quasi-linear kinematic $\eta_{yx}$ by approximately a factor of 2 at $\mathrm{Rm}=100$,
meaning the presence of magnetic fluctuations could change the sign
of $\eta_{yx}$ and excite a coherent large-scale dynamo once $b_{rms}\gtrsim u_{rms}/2$. This
prediction is not far off the observed transition at  $\bm{\sigma}_{\bm{b}}\approx0.2 \bm{\sigma}$ in Fig.~\ref{fig:Low-Rm-magnetic driven},
with the discrepancy presumably arising due to inaccuracies in the 
quasi-linear approximation, as well as the additional presence of an incoherent dynamo mechanism.
Note that $\eta_{xx}$ in the magnetic 
case is much smaller than the kinematic value and will cause only 
a very small (probably unnoticeable) change  to $\eta_{xx}$ unless $u_{rms}\ll b_{rms}$. 
This is basically in agreement with the well-known result that magnetic fluctuations
do not significantly quench the turbulent resistivity. [In analytic SOCA calculations with the 
shear added perturbatively \citep{Analytic} the contribution 
of $b_{rms}$ to $\eta_{xx}$ is exactly zero]. 

Overall, we see that results of Fig.~\ref{fig:b Transport-coefficients} agree 
well with our conclusions from earlier in the section and from Figs.~\ref{fig:mag comparison NRot}, \ref{fig:mag comparison wRot}, and \ref{fig:Low-Rm-magnetic driven}. Magnetic fluctuations
in the presence of shear cause a significant negative contribution to $\eta_{yx}$, which 
can overwhelm (or enhance in the presence of rotation) the kinematic coefficient. 
Thus, with sufficiently strong magnetic fluctuations, a non-helical coherent large-scale 
dynamo is possible through the magnetic shear-current effect.

\section{Discussion and conclusions}\label{Conclusions}

The main purpose of this work has been to propose and explore numerically a novel possibility 
for large-scale magnetic field generation in turbulent plasmas -- the magnetic shear-current effect. 
The basic idea is that in the presence of large scale velocity shear, small-scale magnetic fluctuations
produce an off-diagonal turbulent resistivity ($\eta_{yx}$) with the correct sign 
to cause mean-field dynamo instability when coupled with the shear. 
This is the magnetic analogue of the controversial shear-current
effect \citep{Rogachevskii:2003gg,Rogachevskii:2004cx} and the $\bm{\Omega}\times\bm{J}$ (or Rädler) effect (in
the presence of shear). Importantly, this effect opens the possibility of
the saturated small-scale dynamo \emph{driving} the large-scale dynamo, in stark contrast 
to standard $\alpha$-quenching ideas where the small-scale dynamo
is harmful to mean-field growth. Reassuringly -- and unlike  the kinematic shear-current effect --
the sign of the magnetic effect agrees between analytic SOCA calculations \citep{Analytic}, the 
$\tau$-approximation \citep{Rogachevskii:2004cx}, and quasi-linear theory (\citealt{SINGH:2011ho}; magnetic results 
presented here). 
In addition, all three closure methods agree that the magnetic effect is substantially
larger than the kinematic effect (for similar fluctuation levels $b_{rms}\sim u_{rms}$),
and perturbative MRI shearing wave calculations \citep{Lesur:2008fn}
have also found similar results.
We hope that this agreement speaks to the robustness of the effect in comparison to its kinematic cousin,
at both high and low Reynolds numbers.

In addition to the magnetic dynamo, we have presented results concerning the kinematic shear dynamo, as
studied previously by a number of authors (e.g., \citealt{Rogachevskii:2003gg,Yousef:2008ix,Brandenburg:2008bc,Singh:2013va}).
Our primary result is the qualitative (and quantitative) change in the mean-field dynamo that
occurs due to the addition of rotation. This 
is caused by the well-known $\bm{\Omega}\times\bm{J}$ (or Rädler) effect \citep{Krause:1980vr},  
which for anticyclonic rotation will cause the off-diagonal resistivity $\eta_{yx}$ to have the 
required sign for a mean-field dynamo \citep{Moffatt:1982ta}.
We have seen in a variety of examples how this can cause a change in the mean-field dynamo
from being completely driven by fluctuations in $\alpha$ (the stochastic-$\alpha$ effect), to being at least
partially driven by the off-diagonal turbulent resistivity. The change is observable both qualitatively, in the 
spatiotemporal evolution of $B_{y}$, and quantitatively, in an increase in the dynamo growth rate.

This paper has focused on the dynamo at low Reynolds numbers, similar to that studied by 
\citet{Yousef:2008ix,Yousef:2008ie}. This choice has the advantage of both removing the complications
of small-scale dynamo from the problem, and enabling the use of the quasi-linear approximation 
\citep{Sridhar:2009jg,Squire:2015fk} with some degree of accuracy. The former advantage allows clean and straightforward 
separation of  kinematic and magnetic effects, while the latter enables the use of statistical simulation techniques (CE2) that  
make the differences between incoherent and coherent dynamos particularly transparent. 
Nonetheless, precisely by enabling these simplifications, the low Reynolds number case is also less interesting.
In particular, the magnetic fluctuations cannot arise self-consistently through the small-scale dynamo, which 
is far more natural than a direct forcing of the induction equation (except perhaps in the presence of 
magnetic instabilities such as the MRI). 
To rectify this, in a recent paper \citep{HighRm} we give numerical results 
that illustrate that the magnetic fluctuations \emph{arising from the small-scale dynamo}
can indeed cause a coherent large scale dynamo through $\eta_{yx}$.
 
Given the historical controversy surrounding
some aspects of the shear dynamo, we feel it helpful to give a brief survey of the relationship to several previous works. 
As mentioned in the main text, our results here on the kinematic dynamo agree very nicely with numerical results
in \citet{Yousef:2008ix,Yousef:2008ie}. In particular, our conclusion that rotation fundamentally changes the
shear dynamo is nicely supported by \citet{Yousef:2008ix} Fig.~5, and can even be observed in the spatiotemporal plots of their Fig.~4. We also find basic agreement with 
the quantitative results of \citet{Brandenburg:2008bc}, 
for instance, the transport coefficient calculations showing $\eta_{yx}>0$, since these are carried out kinematically (neglecting 
the Lorentz force). However, we tentatively propose a different interpretation of their Fig.~8 (and possibly Fig.~7), whereby the 
{magnetic} shear-current effect is acting to drive the observed mean-field dynamo coherently (note the high Pm, which should 
lead to strong magnetic fluctuations). In support of this we note the very coherent appearance of the 
dynamo, as well as the near cyclic behavior in the saturation phase (cf.~Fig.~\ref{fig:Low-Rm-magnetic driven}). 
Of course, more work is needed to assess this possibility more thoroughly. Similarly, the $\mathrm{Rm}>1$ simulations
of \citet{Singh:2013va} (Figs.~6-8) may permit a similar explanation, although it is unclear whether there is truly a small-scale dynamo here.
Finally, we mention again the analytic work of \citet{Rogachevskii:2004cx}, where the magnetic shear-current effect is derived within the 
$\tau$-approximation, although the authors do not comment on the result extensively. Specifically, it is clear from their Fig.~3 that
the magnetic effect is far stronger (when the mean-field is zero) than the kinematic effect, in broad agreement with
our results in this work and \citet{Analytic}.

Of course, since this work has explored only the low Reynolds number
regimes, a variety of future studies will be important. While we have 
illustrated that the magnetic shear-current effect can arise 
from the small-scale dynamo in \citet{HighRm}, more work will be
needed to more precisely assess regimes in which the effect may prevail. 
Of particular interest will be the interaction of the effect with 
magnetic helicity conservation arguments. This has been explored
analytically and using quenching models in \citet{Rogachevskii:2006hy} (see
also the appendix of \citealt{Brandenburg:2008bc}), but
more numerical studies would be needed before any definite 
conclusions can be drawn. It would also be interesting to explore 
the relevance of the magnetic shear-current effect in flows with helicity and
a deterministic $\alpha$ effect. Is it possible that the effect could be present, 
perhaps after saturation of the $\alpha\Omega$ dynamo? This may also be
complicated by recent results showing that shear may help 
to enhance helical dynamos by reducing the small-scale field generation \citep{Cattaneo:2014jg,Tobias:2014ek}.

Finally, we note the likely applicability of the magnetic 
shear-current effect to self-sustaining magnetorotational turbulence, where 
magnetic fluctuations are often substantially stronger than velocity fluctuations. 
With the confluence of magnetic fluctuations and anti-cyclonic rotation, it seems reasonable 
to surmise that the magnetic shear-current effect should be important.
Dynamo cycles observed in unstratified magnetorotational turbulence bear some resemblance
to the quasi-periodic behavior the saturated state of the magnetic dynamo (Fig.~\ref{fig:Low-Rm-magnetic driven}), 
and it has been concluded previously that the dynamo arises through a negative $\eta_{yx}$ \citep{Lesur:2008fn,Lesur:2008cv}.
In addition, some of the most solid evidence for the effect's importance
comes from the CE2 simulations in \citet{Squire:2015fk}. Here, since the kinematic 
effect is far too weak and incoherent effects are excluded, the magnetic shear-current effect is the \emph{only possible} 
mechanism to drive the dynamo. The agreement between the saturation of the dynamo 
in CE2 and nonlinear self-sustaining MRI turbulence simulations, in particular through the $\mathrm{Pm}$ dependence,
provides solid evidence that the MRI dynamo is indeed driven by the magnetic shear-current dynamo 
studied in this work. 
Interactions
between this effect and the $\alpha$-effect due to vertical stratification \citep{Gressel:2010dj} may help 
to provide simple mean-field models that could be helpful in observationally useful disk models.

Whatever the outcome of the variety of questions proposed in the 
previous paragraphs, given the generic presence of velocity shear flows and 
magnetic fluctuations in astrophysical plasmas, it seems likely that the proposed 
effects should find application across a wide variety of objects and phenomena.



\appendix

\section{Stochastic-$\alpha$ shear dynamos: some notes on previously proposed mechanisms}\label{app:Sa dynamos}
There has been a wide variety of literature on stochastic-$\alpha$ dynamos in shear flows. Here we consider the relationship
between a number of these works, and explain some fundamental differences that 
would have important consequences for their observation in simulations. We feel 
that this discussion is suitable for presentation in this work, since our 
primary purpose has been to propose an alternative to the stochastic-$\alpha$ mechanism. 

At least two fundamentally different dynamo mechanisms are possible from fluctuations in the $\alpha$ effect with zero mean. 
The first, which has been explored for a variety of perspectives in  \citet{Vishniac:1997jo,Proctor:2007js,
Brandenburg:2008bc,Bushby:2010hy,Heinemann:2011gt,Richardson:2012gf,Mitra:2012iv,
McWilliams:2012du}, has the property (discussed in Sec.~\ref{sec:Shear-dynamos}) that $\left<\bm{B}(t)\right>$ 
decays in time, and only $\langle\bm{B}^{2}\rangle$ undergoes exponential instability. 
We will term this the \emph{incoherent stochastic-$\alpha$ mechanism}.
(We remind the reader that $\left<\cdot\right>$ refers to an ensemble average, while $\overline{\cdot}$ 
refers to the mean-field average.) The second mechanism, which is in essence the Kraichnan-Moffat 
dynamo \citep{Kraichnan:1976gi,Moffatt:1978tc}, has been explored in the context of shear flows in
\citet{Silantev:2000tb,Sridhar:2014he}, and does exhibit growth of $\left<\bm{B}(t)\right>$. 
We shall term this the \emph{coherent stochastic-$\alpha$ mechanism}.
Since part of our argument for the prevalence of coherent dynamo in some of our 
numerical experiments has centered on the requirements on mean-field evolution imposed by $\left<\bm{B}(t)\right>=0$, 
it seems worth explaining in more detail the coherent stochastic-$\alpha$ mechanism and its relation to the incoherent variety. 

In its absolute simplest form, the dynamo in  \citet{Kraichnan:1976gi,Sridhar:2014he}  can be
described as resulting from 
\begin{equation}
\partial_{t}\bm{B} = \nabla \times\left[\alpha(\bm{x},t)\bm{B}\right]+\eta_{T}\nabla^{2}\bm{B}\label{eq:KMinit},
\end{equation}
where $\alpha(\bm{x},t)$ is a spatiotemporal fluctuating $\alpha$-effect, assumed to arise from 
smaller scale fluctuations, and $\eta_{T}$ is the turbulent resistivity. 
One then specifies that $\langle \alpha \rangle=0$, $ \langle\alpha(\bm{x},t)\alpha(\bm{x}',t')\rangle=2\mathcal{A}(\bm{x},\bm{x}')D(t,t')$,
and forms the equation for  $\langle\bm{B}\rangle$
\begin{equation}
\partial_{t}\langle\bm{B}\rangle = \nabla \times\left(\bm{V}_{M}\times \langle \bm{B}\rangle \right)+\eta_{K}\nabla^{2}\langle \bm{B}\rangle \label{eq:KMdynamo},
\end{equation}
where $\eta_{K}\equiv \eta_{T}-\mathcal{A}(0)$, and $\bm{V}_{M}\equiv \int_{0}^{\infty}d\tau \langle \alpha(\bm{x},\tau)\nabla \alpha(\bm{x},0) \rangle$.
For sufficiently strong $\alpha$ fluctuations, instability arises for $\langle\bm{B} \rangle$,
because $\eta_{K}$ becomes negative. Note that for such an instability 
the \emph{smallest} scales of the mean-field grow the fastest. \citet{Sridhar:2014he} give a variety of interesting
extensions to this model, including the effects of non-zero $\alpha$ correlation time $\tau_{\alpha}$, and shear 
(which changes the dynamo only if $\tau_{\alpha} \neq 0$).

Why is it that this dynamo seems to be mean-field in the true sense -- $\langle\bm{B}\rangle$ grows exponentially -- while this
is not true for the incoherent stochastic-$\alpha$ dynamo? This question is 
important for understanding the shear dynamo, since a dynamo arising though 
this coherent stochastic-$\alpha$ mechanism will have very different properties. 
While it seems that all previous treatments 
of this dynamo have considered a spatiotemporal fluctuating $\alpha$ coefficient, this
is not the fundamental difference. In particular, if we simply assert that $\alpha(\bm{x},t)=\alpha(t)$ the
dynamo can still exist with $\eta_{K}\equiv \eta_{T}-\overline{\alpha^{2}}/2$, although $\bm{V}_{M}=0$.
The answer to this question is given in \citet{Mitra:2012iv} Sec.~3.3, where they examine the effects of mutual correlations between
$\alpha$ coefficients. In particular (now considering specifically a horizontal mean-field average such that we have only a 2-D system),  they find that in the presence 
of mutual \emph{correlations} between $\alpha$ coefficients,
\begin{equation}
\left<\alpha_{ij}(t)\alpha_{kl}(t')\right> = \mathcal{D}_{kl}^{ij}\delta(t-t'),
\end{equation}
the ensemble averaged mean-field $\left<\bm{B}\right> = (\langle B_{x}\rangle,\langle B_{y}\rangle )$ satisfies the equation
\begin{equation}
\partial_{t}\left<\bm{B}\right> = 
\left(
\begin{array}{ccc}
  -k^{2}(\eta_{T}+\mathcal{D}^{yx}_{yx}-\mathcal{D}^{yy}_{xx})&    k^{2}(\mathcal{D}^{yy}_{xy}-\mathcal{D}^{yx}_{yy})&   \\
  -S + k^{2}(\mathcal{D}^{xx}_{yx}-\mathcal{D}^{xy}_{xx})&   -k^{2}(\eta_{T}+\mathcal{D}^{xy}_{xy}-\mathcal{D}^{xx}_{yy})&   
\end{array}
\right)\left<\bm{B}\right>.\label{eq:MitraKM}
\end{equation}
Evidently, from Eq.~\eqref{eq:KMinit}, in the coherent stochastic-$\alpha$ mechanism, $\alpha_{xx}(t)=\alpha_{yy}(t)$, 
while $\alpha_{yx}(t)=\alpha_{xy}(t)=0$. This implies $\mathcal{D}^{xx}_{yy}=\mathcal{D}^{xx}_{xx}=\mathcal{D}^{yy}_{yy}=\langle \alpha^{2} \rangle /2$,
while all other $\mathcal{D}^{ij}_{kl}$ vanish. Thus, one obtains exactly the same instability from Eq.~\eqref{eq:MitraKM}, since 
$\eta_{T}-\mathcal{D}^{yy}_{xx}$ can be negative. 

We thus see that the coherent stochastic-$\alpha$ mechanism requires the rather
specific situation of \emph{strong diagonal} $\alpha$ fluctuations, but \emph{very weak
off-diagonal} $\alpha$ fluctuations (since $\mathcal{D}^{yx}_{yx} = \langle \alpha_{yx}^{2}\rangle/2$, and similarly for $\alpha_{xy}$).
While the exact result Eq.~\eqref{eq:MitraKM} is only valid for $\alpha$ with no spatial
dependence, it seems almost certain that similar conclusions will hold if spatial variation is also included. 
Is it realistic for the correlation between $\alpha_{xx}$ and  $\alpha_{yy}$ to greatly exceed the 
fluctuations in $\alpha_{yx}$ and  $\alpha_{xy}$ (their difference 
must also overcome $\eta_{T}$)? Possibly, for instance if the fluctuations in $\alpha_{ij}$
arose purely from fluctuations in small-scale helicity, but this situation
strikes us as unlikely. In any case, it seems that more work, both numerical 
and analytic (e.g., inclusion of $\alpha_{yx}$ and $\alpha_{xy}$ in the much more thorough calculations of \citealt{Sridhar:2014he}), would be needed to 
thoroughly assess this possibility.
Overall, the confluence of factors against the coherent stochastic-$\alpha$ 
dynamo -- the requirement for very strong $\alpha$ fluctuations, the significantly adverse effect of off-diagonal $\alpha$, and
the fact that one would observe a mean-field that grows much faster on the smallest
scales -- leads us to conclude that this mechanism has probably not been observed in previous 
numerical experiments on shear dynamo.

\end{document}